\documentclass[aps,prb,showpacs,tightenlines,twocolumn,secnumarabic,nofootinbib,nobibnotes,superscriptaddress]{revtex4-1}
\usepackage{amsmath,amssymb,amsfonts,bm}
\usepackage{graphicx}
\usepackage{epstopdf}
\usepackage{dcolumn}
\usepackage[colorlinks=true,linkcolor=blue,citecolor=blue, urlcolor=blue]{hyperref}
\begin{document}


\title{Magnetic impurity in a Weyl semimetal}

\author{Jin-Hua Sun}
\affiliation{Department of Physics, Zhejiang University, Hangzhou 310027, China}
\affiliation{Collaborative Innovation Center of Advanced Microstructures, Nanjing 210093,
China}

\author{Dong-Hui Xu}
\affiliation{Department of Physics, Hong Kong University of Science and Technology, Clearwater Bay, Kowloon, Hong Kong, China}

\author{Fu-Chun Zhang}
\affiliation{Department of Physics, Zhejiang University, Hangzhou 310027, China}
\affiliation{Collaborative Innovation Center of Advanced Microstructures, Nanjing 210093,
China}

\author{Yi Zhou}
\affiliation{Department of Physics, Zhejiang University, Hangzhou 310027, China}
\affiliation{Collaborative Innovation Center of Advanced Microstructures, Nanjing 210093,
China}

\date{\today}

\begin{abstract}
We utilize the variational method to study the Kondo screening of a spin-$1/2$ magnetic impurity in a three-dimensional (3D) Weyl semimetal with two Weyl nodes along the $k_z$-axis.  The model reduces to a 3D Dirac semimetal when the separation of the two Weyl nodes vanishes.  When the chemical potential lies at the nodal point, $\mu=0$, the impurity spin is screened only if the coupling between the impurity and the conduction electron exceeds a critical value.  For finite but small $\mu$, the impurity spin is weakly bound due to the low density of state, which is proportional to $\mu^2$, contrary to that in a 2D Dirac metal such as graphene and 2D helical metal where the density of states is proportional to $|\mu|$.  The spin-spin correlation function $J_{uv}(\mathbf{r})$ between the spin $v$-component of the magnetic impurity at the origin and the spin $u$-component of a conduction electron at spatial point $\mathbf{r}$, is found to be strongly anisotropic due to the spin-orbit coupling, and it decays in the power-law. The main difference of the Kondo screening in 3D Weyl semimetals and in Dirac semimetals is in the spin $x$- ($y$-) component of the correlation function in the spatial direction of the $z$-axis.
\end{abstract}


\maketitle


\section{Introduction}
  Two-dimensional (2D) Dirac fermions have been proposed and observed in graphene\cite{Graphene2004,rmp,grapheneinteraction} and in surface states of 3D topological insulators (TIs) \cite{Moore2010,Kane2010,Zhang2011}. 3D Dirac semimetal represents a new state of quantum matter, which can host 3D Dirac fermions in the bulk\cite{Wang2013}. The stable 3D Dirac semimetals have been realized experimentally in $\mathrm{ Na_3Bi}$ compounds \cite{liu2014} and $\mathrm{Cd_3As_2}$ crystals \cite{LiuZK2014,neupane2014}, where the Dirac points are stabilized by crystalline symmetry, and the Dirac nodes are degenerate. If the inversion ($\mathcal{P}$) or time-reversal ($\mathcal{T}$) symmetry is broken, each Dirac node splits into two Weyl nodes resulting in Weyl semimetals \cite{Wan2011,Burkov2011,Vazifeh2013}.
Weyl semimetals show interesting physics such as Fermi arc surface states and chiral anomaly. Recently, Weyl semimetals have attracted much attention because a new TaAs family of Weyl semimetals has been predicted in theories \cite{Weng2015, Huang2015} and subsequently observed in experiments
\cite{Xu2015,Lv2015,Xu20152,Zhang2015,Yang2015,Wang2015,HuangXC2015,Lv2015724}.

\begin{figure}[t]
\begin{center}
\includegraphics[scale=0.46, bb=160 20 370 360]{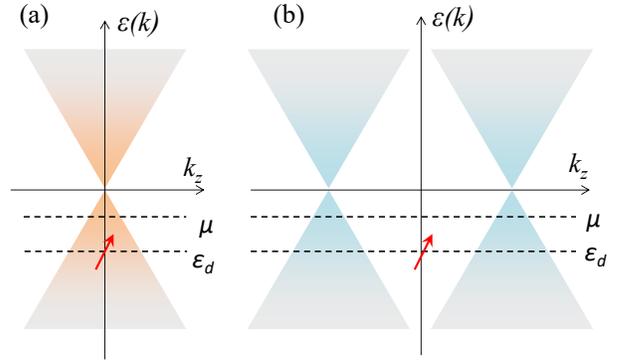}
\end{center}
\caption{(Color online). Schematics of the dispersion relation of (a) Dirac semimetals and (b) Weyl semimetals along the $k_z$-axis. The Dirac cones are located at $k_z=0$ in the Dirac semimetals, and the Weyl nodes are located at $k_z=\pm b/(2\lambda)$ in the Weyl semimetals. $\mu$ is the chemical potential and $\varepsilon_d$ is the energy level of the Anderson impurity.} \label{1_Band.eps}
\end{figure}

 A single magnetic impurity \cite{anderson1961} in a conventional metal is well described by the Anderson impurity model.  This model or the Kondo problem \cite{kondo1964} has been widely studied by using various methods \cite{Krishna1980,tsvelick1984,Andrei1984,Zhang1983,Coleman1984,
 read1983,Kuramoto1983,Gunnarsson1983,affleck1990}. The impurity spin-$1/2$ is fully screened, and the correlation between the impurity spin and a conduction electron of distance $r$ is of a power law decay  $1/r^{d}$ if $r<\xi_K$, and $1/r^{d+1}$ if $r>\xi_K$, with $\xi_K$ the Kondo coherence length and $d$ the dimensionality of the host metal\cite{ishii1978,Barzykin1998,Borda2007}.  Graphene has a peculiar electron structure, where the density of states (DOS) vanishes at a charge neutral point or at half-filling.
 The property of a magnetic impurity in half-filled graphene  falls into the category of pseudo-gap Kondo problem\cite{Gonzalez1998,Fritz2004,Vojta2004}
 which has been studied constantly using various methods including numerical renormalization group (NRG).
 In single layer graphene, the full screening of a magnetic impurity requires a finite strength of the hybridization between the impurity and Dirac electrons, and the spin-spin correlation between the impurity and conduction electron decays with $1/r^3$ power law for large $r$ \cite{shirakawa2014}. Recently the Kondo effect in 3D Dirac and Weyl systems was studied using the NRG method, and it was found that the magnetic impurity shows a diverse range of Kondo physics, depending on the DOS of the host system and the symmetries broken by perturbations\cite{Mitchell2015}.
The spin-spin correlation in the spin-orbit coupled systems may be interesting since the coupling between the spin and the momentum results in anisotropy in both the spin and spatial spaces.

The purpose of this paper is to investigate the properties of spin-spin correlation in Dirac and Weyl semimetals. We study the binding energy and the various components of the spin-spin correlation and we illustrate the similarity or differences between the Dirac and Weyl semimetals. The variational method we apply has been used to study the ground state of the Kondo problem in conventional metals \cite{Gunnarsson1983,Varma1976}, antiferromagnets \cite{Aji2008} and 2D helical metals \cite{Feng2010}.

The paper is organized as follows. We present the model and dispersion relation in Sec.
\ref{model}. In Sec. \ref{binding}, we apply the variational method to study the binding energy. In Sec. \ref{correlation}, we investigate the spin-spin correlation between the magnetic impurity and the conduction electrons in Dirac or Weyl semimetals, and we compare the results in these two systems.
Finally, the discussions and conclusions are given in
Sec. \ref{conclusion}.

\section{Anderson Model Hamiltonian}\label{model}
We utilize the Anderson impurity model to study the Kondo screening of a spin-1/2 magnetic impurity in a 3D Dirac or Weyl semimetal. The model Hamiltonian contains three parts:  the kinetic energy $H_0$ of the Dirac or Weyl semimetal, the impurity Hamiltonian $H_d$, and the hybridization between the impurity and the Dirac or Weyl semimetal $H_V$.  The Hamiltonian reads
\begin{equation}
\begin{aligned}
H=H_0+H_{d}+H_{V}.
\end{aligned}
\end{equation}
Here the kinetic energy part is given by
\begin{equation}
\begin{aligned}
H_0&=\sum_{\mathbf{k}}\Psi_\mathbf{k}^{\dagger} [2\lambda \sigma_z (\mathbf{S}\times\mathbf{k})\cdot \hat{z}+2\lambda_z \sigma_y k_z +\sigma_x M_\mathbf{k} \\
&+ b S_z-\mu]\Psi_\mathbf{k}, \\
&=\sum_{\mathbf{k}} \Psi_\mathbf{k}^{\dagger}
\left(
\begin{array}{cccc}
b-\mu            & 2\mathrm{i}\lambda k_{-}  & -2\mathrm{i}\lambda k_z & 0\\
-2\mathrm{i}\lambda k_{+} & -b-\mu         & 0              & -2\mathrm{i}\lambda k_z\\
2\mathrm{i}\lambda k_z& 0          & b-\mu              & -2\mathrm{i}\lambda k_{-} \\
0            & 2\mathrm{i}\lambda k_z & 2\mathrm{i}\lambda k_{+} & -b-\mu
\end{array}
\right)\Psi_\mathbf{k},
\end{aligned}
\end{equation}
where $\mathbf{k}=(k_x, k_y, k_z)$ and $k_{\pm}=k_x\pm \mathrm{i} k_y$.
$\mathbf{\sigma}$, $\mathbf{S}$ are Pauli matrices in orbital and spin spaces, respectively. The basis vectors of the bulk states are chosen as $\Psi_\mathbf{k}=(a_{\mathbf{k}\uparrow}, a_{\mathbf{k}\downarrow}, b_{\mathbf{k}\uparrow}, b_{\mathbf{k}\downarrow})^{T}$, where a and b are orbital indices. $M_\mathbf{k}=M-tk^2$, with $M$ the Dirac mass.   For $\lambda_z>0$ and $M<0$, $H_0$ describes a strong topological insulators with $Z_2$ index (1; 000). Here we shall consider the case with $M=0$ while the bulk gap closes at $\mathbf{k}=0$ and a transition between a topologically nontrivial phase and a trivial phase occurs.
For simplicity, we shall assume that $\lambda_z=\lambda$ eliminates the extra anisotropy, and we set $M_\mathbf{k}=0$ since the nonzero $M_\mathbf{k}$ merely modifies the high energy dispersion which has a minor influence on our study. At $b=0$, the $H_0$ describes the Dirac semimetal.  At $b \neq 0$, it describes a Weyl semimetal with broken time-reversal ($\mathcal{T}$) symmetry.\cite{FuLiang2010,Zhang2011,Vazifeh2013}
In a Dirac semimetal, the Dirac points are doubly degenerate [see Fig. \ref{1_Band.eps} (a)], while in a Weyl semimetal each Dirac point splits into two Weyl nodes due to the broken $\mathcal{T}$ or parity symmetry. We note that the Weyl semimetal studied in the present paper corresponds to the $\mathcal{T}$-broken case.  For $b>0$, the Weyl nodes are located at the points $(0,0,\pm b/2\lambda)$ on the $k_z$-axis as shown in Fig. \ref{1_Band.eps} (b).

The local magnetic impurity Hamiltonian can be written as
\begin{equation}
\begin{aligned}
H_d=(\varepsilon_d-\mu)(d^{\dagger}_{\uparrow}d_{\uparrow}+d^{\dagger}_{\downarrow}d_{\downarrow})+Un_{d\uparrow}n_{d\downarrow}.
\end{aligned}
\end{equation}
$d_{\uparrow (\downarrow)}^{\dagger}$ and $d_{\uparrow (\downarrow)}$ are the creation and annihilation operators of the spin-up (spin-down) state on the impurity site, and $\varepsilon_d$ and $U$ are the impurity energy level and on-site Coulomb repulsion, respectively.

Finally, the hybridization between the magnetic impurity and the host material is described by
\begin{equation}
\begin{aligned}
H_{V}=\sum_{\mathbf{k}} d_\mathbf{k}^{\dagger} \hat{V} \Psi_\mathbf{k}.
\end{aligned}
\end{equation}
$d_\mathbf{k}=(d_{\mathbf{k}\uparrow}, d_{\mathbf{k}\downarrow})^T$, where $d_{\mathbf{k}s}$ ($d_{\mathbf{k}s}^{\dagger}$) is the impurity annihilation (creation) operators in the momentum space. $\hat{V}$ is the hybridization strength and we assume the magnetic impurity is equally coupled to the four bands in the semimetal,
\begin{equation}
\begin{aligned}
\hat{V}=
\left(
\begin{array}{cccc}
V_k & 0    & V_k  & 0\\
0   & V_k  & 0    & V_k
\end{array}
\right).
\end{aligned}
\end{equation}

The single particle eigen-energy, $\varepsilon_j(\mathbf{k})$, with $j=1,2,3,4$ can be obtained by diagonalizing the non-interacting Hamiltonian $H_0$, and it is given by
\begin{equation}
  \begin{aligned}
  \varepsilon_1(\mathbf{k})&=- \sqrt{4\lambda^2(k_x^2+k_y^2)+(b- 2 \lambda k_z)^2}-\mu,\\
  \varepsilon_2(\mathbf{k})&=- \sqrt{4\lambda^2(k_x^2+k_y^2)+(b+ 2 \lambda k_z)^2}-\mu,\\
  \varepsilon_3(\mathbf{k})&= \sqrt{4\lambda^2(k_x^2+k_y^2)+(b- 2 \lambda k_z)^2}-\mu,\\
  \varepsilon_4(\mathbf{k})&= \sqrt{4\lambda^2(k_x^2+k_y^2)+(b+ 2 \lambda k_z)^2}-\mu,\\
  \end{aligned}
  \end{equation}
where $b=0$ and $b\neq 0$ correspond to the Dirac and Weyl semimetals, respectively. The corresponding eigenstates are given by
\begin{equation}
\begin{aligned}
&\gamma_{\mathbf{k}1}=\frac{1}{\sqrt{C_1}}\{ \phi_{1\mathbf{k}} a_{\mathbf{k}\uparrow} + 2\mathrm{i}\lambda k_{-} a_{\mathbf{k}\downarrow} +\mathrm{i}\phi_{1\mathbf{k}} b_{\mathbf{k}\uparrow} +2\lambda k_{-} b_{\mathbf{k}\downarrow}\},\\
&\gamma_{\mathbf{k}2}=\frac{1}{\sqrt{C_2}}\{ -\phi_{2\mathbf{k}} a_{\mathbf{k}\uparrow} - 2\mathrm{i}\lambda k_{-} a_{\mathbf{k}\downarrow} +\mathrm{i}\phi_{2\mathbf{k}} b_{\mathbf{k}\uparrow} +2\lambda k_{-} b_{\mathbf{k}\downarrow}\},\\
&\gamma_{\mathbf{k}3}=\frac{1}{\sqrt{C_3}}\{ \phi_{3\mathbf{k}} a_{\mathbf{k}\uparrow} + 2\mathrm{i}\lambda k_{-} a_{\mathbf{k}\downarrow} +\mathrm{i}\phi_{3\mathbf{k}} b_{\mathbf{k}\uparrow} +2\lambda k_{-} b_{\mathbf{k}\downarrow}\},\\
&\gamma_{\mathbf{k}4}=\frac{1}{\sqrt{C_4}}\{ -\phi_{4\mathbf{k}} a_{\mathbf{k}\uparrow} - 2\mathrm{i}\lambda k_{-} a_{\mathbf{k}\downarrow} +\mathrm{i}\phi_{4\mathbf{k}} b_{\mathbf{k}\uparrow} +2\lambda k_{-} b_{\mathbf{k}\downarrow}\},  \\
\end{aligned}
\end{equation}
where $C_j$ are the normalization factors.
The parameters $\phi_{l\mathbf{k}}$ ($l=1,2,3,4$) are given by
\begin{equation}
\begin{aligned}
\phi_{1\mathbf{k}}=(b-2\lambda k_z)-m_\mathbf{k}, \ \phi_{2\mathbf{k}}=(b+2\lambda k_z)-n_\mathbf{k}, \\
\phi_{3\mathbf{k}}=(b-2\lambda k_z)+m_\mathbf{k}, \ \phi_{4\mathbf{k}}=(b+2\lambda k_z)+n_\mathbf{k},
\end{aligned}
\end{equation}
where $m_\mathbf{k}(n_\mathbf{k})=\sqrt{4\lambda^2(k_x^2+k_y^2)+(b\mp 2\lambda k_z)}$.
We can rewrite the total Hamiltonian $H$ in the diagonal basis as
\begin{equation}
\begin{aligned}
H=&\sum_{\mathbf{k}j}(\varepsilon_{j}(\mathbf{k})-\mu)\gamma _{\mathbf{k}j}^{\dagger}\gamma _{\mathbf{k}j}+\sum _{\mathbf{k}j}V_\mathbf{k}(\gamma _{\mathbf{k}j}^{\dagger}d_{\mathbf{k}j}+H.C.)\\
  &+(\varepsilon_d-\mu)\sum _{\sigma }d_{\sigma }^{\dagger}d_{\sigma }+U \ d_{\uparrow }^{\dagger}d_{\uparrow }d_{\downarrow }^{\dagger}d_{\downarrow },
\end{aligned}
\end{equation}
where the impurity operators $d_{\mathbf{k}j}$ are given by
\begin{equation}
\begin{aligned}
&d_{\mathbf{k}1}=\frac{(1+\mathrm{i})\phi_{1\mathbf{k}}}{\sqrt{C_1}} d_{\uparrow} + \frac{(1+\mathrm{i})2\lambda k_{-}}{\sqrt{C_1}} d_{\downarrow},\\
&d_{\mathbf{k}2}=\frac{-(1-\mathrm{i})\phi_{2\mathbf{k}}}{\sqrt{C_2}} d_{\uparrow} + \frac{(1-\mathrm{i})2\lambda k_{-}}{\sqrt{C_2}} d_{\downarrow},\\
&d_{\mathbf{k}3}=\frac{(1+\mathrm{i})\phi_{3\mathbf{k}}}{\sqrt{C_3}} d_{\uparrow} + \frac{(1+\mathrm{i})2\lambda k_{-}}{\sqrt{C_3}} d_{\downarrow},\\
&d_{\mathbf{k}4}=\frac{-(1-\mathrm{i})\phi_{4\mathbf{k}}}{\sqrt{C_4}} d_{\uparrow} + \frac{(1-\mathrm{i})2\lambda k_{-}}{\sqrt{C_4}} d_{\downarrow}.\\
\end{aligned}
\end{equation}

\section{The binding energy}\label{binding}
We start from the simplest limit in which the magnetic impurity and the host material are decoupled from each other, namely $H_{V}=0$.
The ground state wavefunction of $H_0$ can be written as
\begin{equation}
\begin{aligned}
|\Psi \rangle_0 =\prod _{\textbf{k}j;\varepsilon _{j}(\mathbf{k})<\mu} \ \gamma _{\mathbf{k}j}{}^{\dagger}|0\rangle ,
\end{aligned}
\end{equation}
where the product runs over all the states inside the Fermi surface.  If $\varepsilon_d<\mu<\varepsilon_d+U$, the impurity is singly occupied with a local moment, and the impurity energy is $\varepsilon_d-\mu$. The total energy of the system is given by the sum of the energies of the host material and of the magnetic impurity,
\begin{equation}
\begin{aligned}
E_0=\varepsilon_d-\mu+\sum_{\mathbf{k}j; \varepsilon _{j}(\mathbf{k})<\mu}\varepsilon_j(\mathbf{k}).
\end{aligned}
\end{equation}
In the above equations, the index $j=1,2,3,4$ denotes the four bands in the Dirac or Weyl semimetal.

\begin{figure}[htpb]\label{4_bindingE.eps}
\begin{center}
\includegraphics[scale=0.4, bb=200 40 480 510]{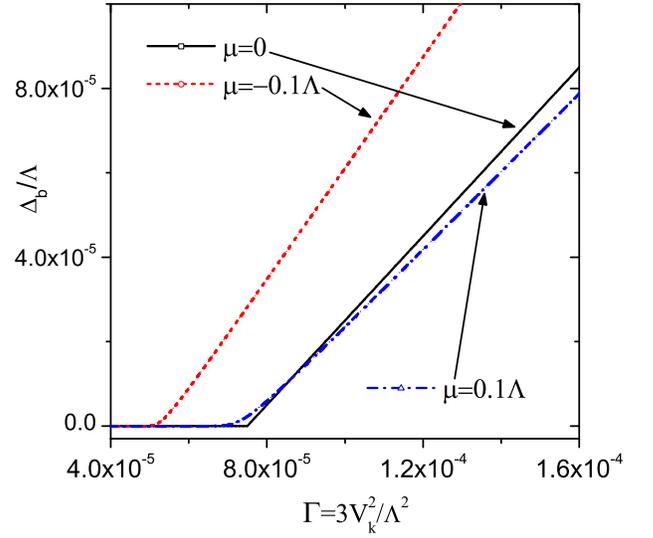}
\end{center}
\caption{(Color online). Calculated binding energy $\Delta_b$ of a magnetic impurity in Dirac or Weyl semimetal as a function of $\Gamma$ for three values of chemical potential $\mu$. $\Lambda$ is the energy cut-off, and $\Gamma=3V_k^2/\Lambda^2$ is the effective hybridization. At $\mu=0$, there is a threshold $\Gamma > \Gamma_c=|\epsilon_d|/\Lambda=7.5\times 10^{-5}$ for the bound state.  For $\mu \neq 0$, $\Delta_b$ is finite, but too small to be seen in the figure for small $\Gamma$.} \label{4_bindingE.eps}
\end{figure}

Now we study the effect of $H_{V}$ which describes the hybridization between the magnetic impurity and the host Weyl semimetal by using a variational method.
We shall assume large-$U$ limit. In this case, we may neglect the doubly occupied impurity states in our trial wavefunction for the ground state, which is given by
\begin{equation}\label{Eq:newwavef}
\begin{aligned}
|\Psi \rangle =(c_0+\sum _{\textbf{k}j}c_{\mathbf{k}j}d_{\mathbf{k}j}^{\dagger}\gamma _{\mathbf{k}j})|0\rangle.
\end{aligned}
\end{equation}
$c_{0}$ and $c_{\mathbf{k}j}$ are variational parameters to be determined by optimizing the ground state energy.
The energy of the Hamiltonian $H$ in the trial state $|\Psi \rangle$ is given by
\begin{equation}\label{Eq:totalE}
\begin{aligned}
E=\frac{\sum_{\mathbf{k}j}[(E_0-\varepsilon_{j}(\mathbf{k})+\mu)c_{kj}^{2}+2V_\mathbf{k} c_0 c_{\mathbf{k}j}+(\varepsilon_{j}(\mathbf{k})-\mu)c_{0}^2]}{c_0^2+\sum_{\mathbf{k}j}c_{\mathbf{k}j}^2},
\end{aligned}
\end{equation}
The variational principle requires
\begin{equation}\label{Eq:totalE}
\begin{aligned}
\partial E/\partial c_0=\partial E/\partial c_{\mathbf{k}j}=0,
\end{aligned}
\end{equation}
from which we obtain two equations below:
\begin{equation}
\begin{aligned}
&Ec_0=\sum_{\mathbf{k}j}[V_\mathbf{k}c_{\mathbf{k}j}+(\varepsilon_{j}(\mathbf{k})-\mu)c_0],\\
&Ec_{\mathbf{k}j}=(E_0-\varepsilon_{j}(\mathbf{k})+\mu)c_{\mathbf{k}j}+V_\mathbf{k}c_0,
\end{aligned}
\end{equation}
which lead to
\begin{equation}\label{Eq:selfConsis}
\begin{aligned}
&(E-\sum_{\mathbf{k}j}(\varepsilon_{j}(\mathbf{k})-\mu))c_0=\sum_{\mathbf{k}j}V_\mathbf{k}c_{\mathbf{k}j},\\
&[E-E_0+(\varepsilon_{j}(\mathbf{k})-\mu)]c_{\mathbf{k}j}=V_\mathbf{k}c_0.
\end{aligned}
\end{equation}
We define the binding energy as $\Delta_b=E_0-E$. If $\Delta_b>0$, then the hybridized state has lower energy than the bare state, so that the hybridized state is more stable.
From Eq. \ref{Eq:selfConsis} we obtain
\begin{equation}\label{Eq:aki}
\begin{aligned}
c_{\mathbf{k}j}=\frac{V_\mathbf{k}c_0}{(\varepsilon_{j}(\mathbf{k})-\mu)-\Delta_b},
\end{aligned}
\end{equation}
and
\begin{equation}
\begin{aligned}
(E-E_0+(\varepsilon_d-\mu))c_0=\sum_{\mathbf{k}j}V_\mathbf{k}c_{\mathbf{k}j}.
\end{aligned}
\end{equation}
We then obtain the self-consistent equation
\begin{equation}\label{Eq:selfconFinal}
\begin{aligned}
(\varepsilon_d-\mu)-\Delta_b=\sum_{\mathbf{k}j}\frac{V_\mathbf{k}^2}{(\varepsilon_{j}(\mathbf{k})-\mu)-\Delta_b}.
\end{aligned}
\end{equation}

By solving Eq. \ref{Eq:selfconFinal} numerically, we obtain the binding energy $\Delta_b$.  From Eq. \ref{Eq:aki} we can calculate $c_{\mathbf{k}j}$ for given $\mathbf{k}$ and $j$.  In the calculations, we introduce an energy cutoff $\Lambda$, and hence the truncation of momentum $k_c=\Lambda/(2\lambda)$.  The summation over momentum in Eq. \ref{Eq:selfconFinal} is then replaced by an integration
$\frac{1}{N}\sum_{\mathbf{k}}\rightarrow \frac{6 \pi^2}{k_{c}^3}\int \frac{\mathbf{d}^3 k}{(2\pi)^3}$. \\
One can see that the binding energy is independent of the host system to be the Dirac or Weyl semimetal . Although the Weyl nodes are located at $k_z=\pm b/2\lambda$ in the Weyl semimetal, the dispersion relation around each Weyl node is exactly the same as those around the Dirac cones, thus the summation over $k$ on the right-hand side of Eq. \ref{Eq:selfconFinal} shall also remain the same.
We define an effective hybridization $\Gamma=\frac{3 V_k^2}{\Lambda^2}$. From Eq. \ref{Eq:selfconFinal} we obtain
\begin{equation}\label{Eq:selfconAna}
\begin{aligned}
(\varepsilon_d-\mu)-\Delta_b+\Gamma[\frac{(\Lambda^2-\mu^2)}{\Lambda}-2\frac{(\Lambda-|\mu|)(\mu+\Delta_b)}{\Lambda}]\\
=-2\Gamma\frac{(\mu+\Delta_b)^2}{\Lambda}\mathrm{ln}\frac{\Lambda+\mu+\Delta_b}{\Delta_b}.
\end{aligned}
\end{equation}
Then in the limit of small $\Gamma$ and $\Gamma\Lambda<\mu-\varepsilon_d+2\Gamma\mu$ we have
\begin{equation}\label{Eq:selfconre}
\begin{aligned}
\Delta_b\approx \Lambda \mathrm{exp} \{ -\frac{\mu-\varepsilon_d-\Gamma\Lambda+2\Gamma\mu}{2\Gamma\mu^2/\Lambda} \}, \ (\mu\neq 0).\\
\end{aligned}
\end{equation}
If $\mu \ne 0$, the hybridization shall always lead to a finite binding energy $\Delta_b>0$. However, if $\mu=0$, Eq. \ref{Eq:selfconAna} gives rise to
\begin{equation}
\begin{aligned}
\varepsilon_d+\Gamma\Lambda=\Delta_b-\frac{2\Gamma}{\Lambda}\Delta_b^2\ln \frac{\Lambda}{\Delta_b}.
\end{aligned}
\end{equation}
In the limit of small $\Gamma$, we obtain $\Delta_b\approx \varepsilon_d+\Gamma\Lambda$.
The density of states in the Dirac or Weyl semimetal vanishes, hence the binding energy $\Delta_b$ has a positive solution only when the effective hybridization $\Gamma$ is above a critical value, $\Gamma>\frac{|\varepsilon_d|}{\Lambda}$.  This is similar to the case of graphene \cite{Sengupta2008,Zhuang2009} and other 2D helical metals \cite{Feng2010}. In the context of Kondo physics, it is so called pseudo-gap Kondo problem.\cite{Gonzalez1998,Fritz2004,Vojta2004}

The self-consistent solution of the binding energy is plotted in Fig. \ref{4_bindingE.eps} as a function of the effective hybridization $\Gamma$. We fix the value of $\mu-\epsilon_d=7.5\times 10^{-5}\Lambda$, and the chemical potential is chosen as $\mu=0$ and $\mu=\pm 0.1\Lambda$.
While $\mu=0$ the system is at half-filling, and we can see that the binding energy $\Delta_b$ is non-zero only if the effective hybridization is greater than the critical value $\Gamma_c=|\epsilon_d|/\Lambda=7.5\times 10^{-5}$.
While $\mu\neq 0$ the binding energy always has a positive value, which can also be seen from the analytical results shown in Eq. \ref{Eq:selfconre}. The DOS $D(E)\propto E^2$ is much smaller than that in graphene near the charge neutral point, so that the screening effect and thus the binding energy are much smaller than those in the graphene case. The asymmetry between the $\mu=\pm0.1\Lambda$ cases is due to the asymmetry of the impurity state between being empty and doubly occupied.

\section{Spin-spin correlation between magnetic impurity and conduction electrons}\label{correlation}
\begin{figure}[t]
\begin{center}
\includegraphics[scale=0.81, bb=140 50 200 500]{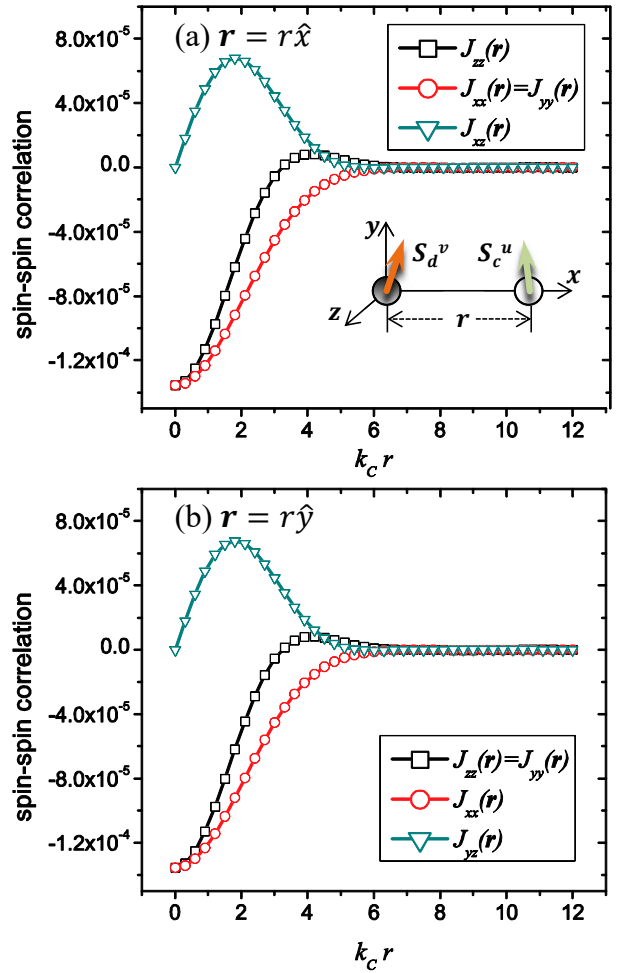}
\end{center}
\caption{(Color online). Spin-spin correlation between magnetic impurity and conduction electron along the $x$-axis (panel a), and along the $y$-axis (panel b). The results are the same for the Dirac or Weyl semimetal.  The inset in panel (a) illustrates spin $S_d^v$ of magnetic impurity at the origin $r=0$ and conduction electron spin $S_c^u$ at a distance $r$ along the $x$-axis.
 The parameters are $\mu=-0.01\Lambda$, $V_k=0.05\Lambda$ and $\Delta=0.05\Lambda$, and the energy cut-off $\Lambda$ is large enough that the value of $\Lambda$ shall not affect the low-energy physics. All the other spin-spin correlations not shown here are zero.} \label{1_sscorr_xy.eps}
\end{figure}

\begin{figure}[htpb]
\begin{center}
\includegraphics[scale=0.46, bb=200 50 385 460]{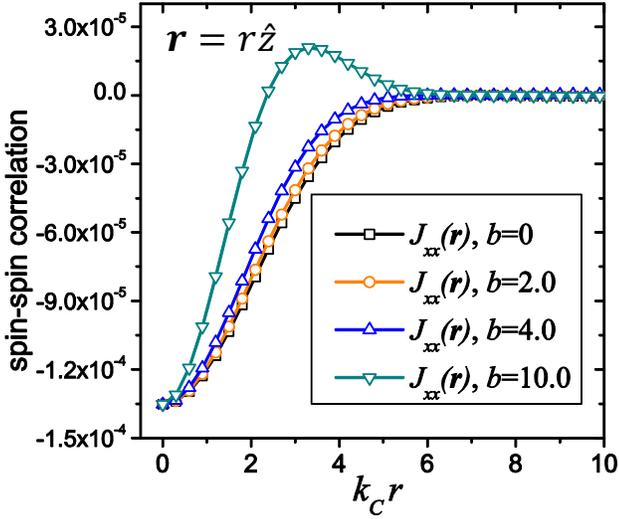}
\end{center}
\caption{(Color online). Spin-spin correlation between a magnetic impurity at the origin and a conduction electron at a distance $r$ along the $z$-axis for Dirac semimetal ($b=0$) and Weyl semimetal of three values of $b$. The two Weyl nodes are at $b/\lambda$ along the $k_z$-axis. Correlation for spins along $z$, $J_{zz}(r)$, is independent of b and $J_{zz}(r)|_b=J_{xx}|_{b=0}=J_{yy}|_{b=0}$, while $J_{xx}(r)$ and $J_{yy}(r)$ are $b$-dependent. The parameters are $\mu=-0.01\Lambda$, $V_k=0.05\Lambda$ and $\Delta=0.05\Lambda$.
 }\label{2_sscorr_zz.eps}
\end{figure}

Now we study the spin-spin correlation between the magnetic impurity $S_d=\frac{1}{2}d^{\dagger}\mathbf{\sigma}d$ and the conduction electron spin $S_c=\frac{1}{2}c^{\dagger}\mathbf{\sigma}c$ in the Dirac or Weyl semimetal.
The impurity-spin\textbf{-}conduction-electron-spin correlation function is evaluated for $\mu\neq 0$ while the binding energy $\Delta_b$ always has a positive value, i.e., there exists Kondo screening.
We assume that the impurity is located at the origin $\mathbf{r}=0$, and we only consider the simplest case in which the spin-spin correlation function is evaluated along the three axes: $x$-, $y$- and $z$-axes. The spin-spin correlation function between the magnetic impurity and the conduction electron is given by
\begin{equation}\label{Eq:sscorr}
\begin{aligned}
J_{uv}(\mathbf{r})=\langle S_{c}^{u}(\mathbf{r})S_{d}^{v}(0)\rangle,
\end{aligned}
\end{equation}
where $u,v=x,y,z$ and $\langle \dots \rangle$ denotes the ground state average.
We start with simple symmetry analysis. As we have mentioned, the Dirac semimetals we study in this paper preserve both $\mathcal{P}$ and $\mathcal{T}$ symmetries, while in Weyl semimetals the $\mathcal{T}$ symmetry is broken due to the displacement of the Dirac cones along the $k_z$-axis. Both the systems have rotational symmetry along the $z$-axis, so we have $J_{uv}(\mathbf{r})=J_{u^\prime v^\prime}(\mathbf{r}^\prime)$ if $u^\prime=\mathcal{R}_z(\beta)u$, $v^\prime=\mathcal{R}_z(\beta)v$, $r^\prime=\mathcal{R}_z(\beta)r$ where $\mathcal{R}_z(\beta)$ is a rotational operator along the $z$-axis.
Actually one can demonstrate that both the systems are unchanged under a combined mirror reflection and time-reversal transformation. We may denote the mirror reflection with respect to the $y-z$ plane as $\mathcal{M}^{yz}$, then we have $\mathcal{TM}^{yz}(x,y,z)=(-x,y,z)$, $\mathcal{TM}^{yz}(k_x,k_y,k_z)=(k_x,-k_y,-k_z)$ and $\mathcal{TM}^{yz}(S_x,S_y,S_z)=(-S_x,S_y,S_z)$. Thus if $\mathbf{r}$ is on the $x$-, $y$- or $z$-axis, we can easily find that only $J_{yz}(r\hat{y})=-J_{zy}(r\hat{y})$ and $J_{xz}(r\hat{x})=-J_{zx}(r\hat{x})$ are nonzero among the off-diagonal terms. We should emphasize that due to the absence of the spin SU(2) symmetry, the off-diagonal terms can generally be non-zero in the coordinate space, but we may only concentrate on the three spatial axes in this present paper.

We can define functions below and use them to simplify the spin-spin correlation function in the coordinate space.
\begin{equation}\label{Eq:defineFunc}
\begin{aligned}
I_j(\mathbf{r})&=\sum_k e^{-i\mathbf{kr}}\frac{\phi_{kj}}{C_j} c_{kj},\\
J_j(\mathbf{r})&=\sum_k e^{-i\mathbf{kr}}\frac{\phi_{kj}}{C_j}2\lambda (-\mathrm{i}k_{+}) c_{kj}, \\
T_j(\mathbf{r})&=\sum_k e^{-i\mathbf{kr}}\frac{\phi_{kj}}{C_j}2\lambda (\mathrm{i}k_{-}) c_{kj},\\
Y_j(\mathbf{r})&=\sum_k e^{-i\mathbf{kr}}c_{kj},\\
Q_j(\mathbf{r})&=\sum_k e^{-i\mathbf{kr}}\frac{4\lambda^2 (k_{x}^{2}+k_{y}^{2})}{C_j}c_{kj}=\frac{1}{2}Y_j-I_j.\\
\end{aligned}
\end{equation}
Here again $j=1,2,3,4$ are used to denote the four bands in the Dirac or Weyl semimetals.
Then the diagonal terms and the non-zero off-diagonal terms of the spin-spin correlation function along the three axes are given by

\begin{widetext}
\begin{equation}\label{Eq:sscorrdia}
\begin{aligned}
J_{zz}(\mathbf{r})&=-\{[|I_1+I_3|^2+|I_2+I_4|^2+|Q_1+Q_3|^2+|J_2+J_4|^2-|J_1+J_3|^2-|J_2+J_4|^2-|T_1+T_3|^2-|T_2+T_4|^2]\},\\
J_{xx}(\mathbf{r})&=-\{2[(J_1+J_3)(T_{2}^*+T_{4}^*)+(J_2+J_4)(T_{1}^*+T_{3}^*)+(I_1+I_3)(Q_2^*+Q_4^*)+(I_2+I_4)(Q_1^*+Q_3^*)]\},\\
J_{yy}(\mathbf{r})&=-\{2[-(J_1+J_3)(T_{2}^*-T_{4}^*)+(J_2+J_4)(T_{1}^*+T_{3}^*)+(I_1+I_3)(Q_2^*+Q_4^*)+(I_2+I_4)(Q_1^*+Q_3^*)]\};\\
J_{xz}(\mathbf{r})&=-2\mathrm{Re}\{[I_1+I_3+\mathrm{i}(I_2+I_4)][T_1^{*}+T_3^{*}-\mathrm{i}(T_2^{*}+T_4^{*})]-[J_1+J_3-\mathrm{i}(J_2+J_4)][Q_1^{*}+Q_3^{*}+\mathrm{i}(Q_2^{*}+Q_4^{*})]\},\\
J_{yz}(\mathbf{r})&=-2\mathrm{Im}\{[I_1+I_3+\mathrm{i}(I_2+I_4)][T_1^{*}+T_3^{*}-\mathrm{i}(T_2^{*}+T_4^{*})]-[J_1+J_3-\mathrm{i}(J_2+J_4)][Q_1^{*}+Q_3^{*}+\mathrm{i}(Q_2^{*}+Q_4^{*})]\}.
\end{aligned}
\end{equation}
\end{widetext}

If we consider the simplest case when $\mu<0$ and $b=0$ then we have $m=2\lambda k$ and $\phi_{1k}=-2\lambda(k+k_z)=-2\lambda k(1+cos\theta)$, and then the correlation writes
\begin{equation}
  \begin{aligned}
J_{zz}\propto |I_1|^2 + |I_2|^2 + |Q_1|^2+|Q_2| ^2.
\end{aligned}
  \end{equation}
We take the first term for example
\begin{equation}\label{Eq:I1_1}
\begin{aligned}
I_1&=\sum_{k}e^{-\mathrm{i}\mathbf{kr}}\frac{\phi_{1k}^2}{C_1}c_{k1}=\sum_ke^{-\mathrm{i}\mathbf{kr}}\frac{\phi_{1k}^2}{-4m\phi_{1k}}c_{k1}\\
   &=\sum_ke^{-\mathrm{i}\mathbf{kr}}\frac{\phi_{1k}}{-4m_k}c_{k1} =\frac{1}{4}\sum_k e^{-\mathrm{i}\mathbf{kr}}(1+\cos \theta)c_{k1}.
\end{aligned}
\end{equation}
After a straightforward integration, we obtain
\begin{equation}\label{Eq:I1_2}
\begin{aligned}
I_1 &\propto \frac{\mathrm{i}}{r} (e^{-\mathrm{i}\frac{\Lambda}{2\lambda}r}-e^{-\mathrm{i}\frac{|\mu|}{2\lambda}r}) \\ &-\frac{\mu+\Delta_b}{2\lambda}e^{\mathrm{i}\frac{(\mu+\Delta_b)r}{2\lambda}}\int_{r\Delta_b/2\lambda}^{r\Lambda/2\lambda}\frac{e^{-\mathrm{i}y}}{y} \mathrm{d}y
\end{aligned}
\end{equation}
By assuming $\mu+\Delta_b\ll \Lambda$, we can ignore the second term and obtain
\begin{equation}\label{Eq:I1_2}
\begin{aligned}
I_1= -\frac{3\mathrm{i} V_k a_0}{4\Lambda }\frac{1}{(k_c r)^2}(e^{-\mathrm{i}\frac{\Lambda}{2\lambda}r}-e^{-\mathrm{i}\frac{|\mu|}{2\lambda}r}),
\end{aligned}
\end{equation}
where $a_0$ is the number defined in Eq. \ref{Eq:newwavef}, and $\Lambda$ is the energy cut-off.
We can easily find that $I_2=I_1$, and $Q_1$ and $Q_2$ also decay with $1/(k_c r)^2$, which indicates that the spin-spin correlation $J_{zz}(r)$
follows a $1/(k_c r)^4$ decay at long distance. Using similar analysis, we can see that while $b=0$ all the other diagonal and off-diagonal components of the spin-spin correlation decay with $1/(k_c r)^4$ in the real space. The results for the Weyl semimetal ($b\ne 0$) are more complicated, which will be discussed later.

In Fig. \ref{1_sscorr_xy.eps}, we show the results of $J_{uv}(r)$ for $r$ along the $x$- and $y$-axis. The values of $J_{uv}(r)$ along these two axes are independent of $b$, hence they are the same for the Dirac and Weyl semimetals.
As shown in Fig. \ref{1_sscorr_xy.eps}, the diagonal terms are all antiferromagnetic at short distance. In Fig. \ref{1_sscorr_xy.eps} (a) for $\mathbf{r}$ along the $x$-axis, we have $J_{xx}(r)=J_{yy}(r)\ne J_{zz}(r)$ and only two of the off-diagonal terms, $J_{xz}(r)=-J_{zx}(r)$, are non-zero. The inset in Fig. \ref{1_sscorr_xy.eps} (a) shows the schematics of the displacement of the magnetic impurity and the conduction electrons. We assume the location of the impurity is at $r=0$ and the conduction electron is on the $x$-axis, and the displacement between them is $r$. $S_d^v$ and $S_c^u$ are used to denote the spins on the magnetic impurity and the conduction electron, respectively.
Similarly in Fig. \ref{1_sscorr_xy.eps} (b) we have $J_{yy}=J_{zz}\ne J_{xx}$ and only one off-diagonal term $J_{yz}$ is non-zero.
The off-diagonal terms reflect the spin-orbit coupling in the Dirac and Weyl semimetals.

In Fig. \ref{2_sscorr_zz.eps}, we show the spin-spin correlation with $\mathbf{r}$ along the $z$-axis, which depends on $b$, hence is different between the Dirac and Weyl semimetals.  Firstly, in the 3D Dirac system, the diagonal terms are all equivalent, $J_{xx}=J_{yy}=J_{zz}$. As the value of $b$ increases, the Dirac cones split and Weyl nodes emerge. One can see that $J_{zz}(r)$ remains the same as $b$ increases. We always have $J_{zz}(r)|_{b=0}=J_{zz}(r)|_{b>0}=J_{xx}(r)|_{b=0}=J_{yy}(r)|_{b=0}$ if $\mathrm{r}$ is along the $z$-axis.
We still have $J_{xx}=J_{yy}$ for $b>0$, and we find that these two terms are modified in the coordinate space as $b$ increases.
If $\mu<0$ and $\mathbf{r}=r\hat{z}$, then from Eq. \ref{Eq:sscorrdia} we obtain
\begin{equation}
\begin{aligned}
J_{xx}(r\hat{z})=-2(J_1T_2^*+J_2T_1^*+I_1Q_2^*+I_2Q_1^*).
\end{aligned}
\end{equation}
where the indices 1 and 2 denote the lower bands whose Weyl nodes are located at $k_z=-b/2\lambda$ and $k_z=b/2\lambda$, respectively. $J_1$ corresponds to the contribution from the lower band whose Weyl node is at $k_z=-b/2\lambda$. If we perform a simple translation along the $k_z$ axis and substitute $k_z$ with $k_z^{\prime}+b/2\lambda$, we obtain $J_1=J_1^{D} e^{-i\frac{b}{2\lambda}r}$ where $J_1^D$ is the results for Dirac semimetal. Similarly, we can see that $T_2^{*}=T_2^{D*}e^{-i\frac{b}{2\lambda}r}$, such that $J_1T_2^*=J_1^{D}T_2^{D*}e^{-i\frac{b}{\lambda}r}$ and $J_2T_1^{*}=J_2^{D}T_1^{D*}e^{i\frac{b}{\lambda}r}$. Given that the two lower bands are degenerate at $b=0$, we obtain
\begin{equation}\label{Eq:Jxx}
\begin{aligned}
J_{xx}(r\hat{z})=J_{xx}^{D}(r\hat{z}) \cos(br/2\lambda),
\end{aligned}
\end{equation}
where $J_{xx}^{D}(r\hat{z})$ is the $x-x$ spin-spin correlation along the $z$-axis in the Dirac semimetal, and $b/2\lambda$ is half of the separation of the two Weyl nodes along the $k_z$-axis. One can easily find that the displacement of the Weyl nodes along the $k_z$-axis will add an extra phase factor to each function defined in Eq. \ref{Eq:defineFunc}, and it will further induce complexity to the oscillation behavior of the spin-spin correlations along the $z$-axis. One can use a similar approach to prove that for the $y-y$ correlation it shall be $J_{yy}(r\hat{z})=J_{yy}^{D}(r\hat{z})\cos(br/2\lambda)$.
Generally in the short distance limit, as shown in Fig. \ref{2_sscorr_zz.eps}, the spin-spin correlation decays faster as $b$ increases.
\begin{figure}[t]
\begin{center}
\includegraphics[scale=0.47, bb=200 50 385 460]{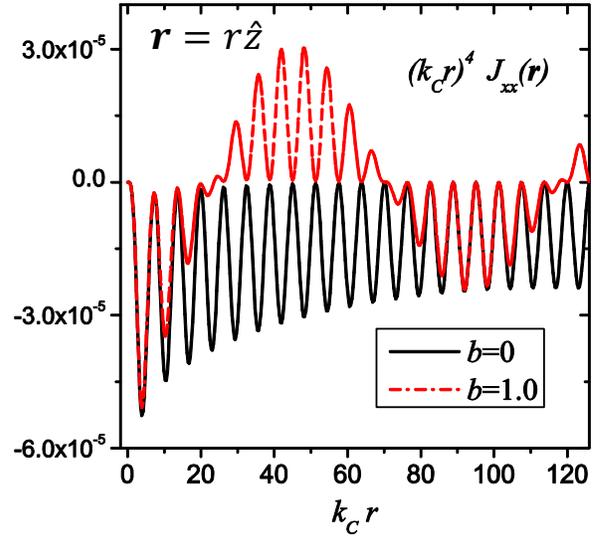}
\end{center}
\caption{(Color online). The product $(k_cr)^4 J_{xx}$ as a function of the dimensionless distance $k_cr$. Here the displacement $\mathbf{r}$ is along the $z$-axis. We use $\mu=-0.01\Lambda$, $V_k=0.05\Lambda$ and $\Delta=0.05\Lambda$. The magnetic impurity is coupled to (a) Dirac semimetals ($b=0$) and (b) Weyl semimetals ($b=1$), respectively.
 }\label{3_decay.eps}
 \end{figure}

Shown in Fig. \ref{3_decay.eps} is the product $(k_cr)^4 J_{xx}(r)$ as a function of the dimensionless distance $k_cr$. We use $\mu=-0.01\Lambda$, $V_k=0.05\Lambda$ and $\Delta=0.05\Lambda$.
 We can see that both the terms for $b=0$ and $b=1.0$ oscillate in coordinate space, and the decay rate is proportional to $(k_cr)^{-4}$.
 This decay rate is consistent with the general results in normal metals that the spin-spin correlation decays with $r^{d+1}$ $(d=3)$ at far displacement.
By carefully examining Eq. \ref{Eq:sscorrdia}, we find that the effect of $b$ is to add an extra phase factor to the spin-spin correlations, i.e., as the phase factor $\cos(br/2\lambda)$
added to the spin-spin correlation $J_{xx}^D(r\hat{z})$
given in Eq. \ref{Eq:Jxx}. The results for different values of $b$ generally have similar contributions, i.e., that the oscillation in the coordinate space is modified by $b$.

\section{DISCUSSION AND CONCLUSION}\label{conclusion}
In conclusion, we have studied spin-$1/2$ Anderson impurity in a Dirac or Weyl semimetal. We apply the variational method to study the problem at the large-$U$ limit. Due to the spin-orbit coupling in these two systems, the spatial spin-spin correlations between the magnetic impurity and the conduction electron are highly anisotropic. The results of the binding energy are quite similar to those in graphene and 2D helical metals, and they are found to be the same for the Dirac and Weyl semimetal. Due to the vanishing DOS at half-filling, there exists a critical value of the effective hybridization strength. We may obtain a positive binding energy only if the hybridization is above the critical value. While the system is particle-hole asymmetric, the DOS at the Fermi surface becomes finite, such that the hybridization always leads to a positive binding energy. However, due to the fact that the DOS near the Weyl points is proportional to $E^2$, the screening effect of the conduction electrons is much weaker than that in graphene, in which $DOS(E)\propto
|E|$.  Therefore, in the weak hybridization limit and at the chemical potential near the Dirac point, the impurity binding energy in the 3D Dirac or Weyl semimetal is much smaller than that in the corresponding graphene system or 2D helical metal.

The spin-spin correlations in both the Dirac and Weyl semimetal are highly asymmetric in coordinate space due to the spin orbit coupling. Although the Kondo temperature of a Dirac or Weyl semimetal is mainly determined by the DOS and not affected by the spin orbit coupling \cite{Rossi2014}, the spin-spin correlation we study here shows rich features spatially.
In general, the diagonal terms decay with $r^{-4}$ for $\mu\ne 0$, which is in agreement with the results in normal 3D metals. The spin-spin correlations are independent of the Dirac or Weyl semimetal, except for the spatial separation between the impurity and conduction electron along  the $z$-axis.  In that case, the momentum $b/2\lambda$ adds an extra phase factor to spin-spin correlation $J_{xx}(r)$ and $J_{yy}(r)$, so that the spin-spin correlation is modified according to $b$.

\section{Acknowledgement}
This work is supported in part by National Basic Research Program of China
(2014CB921201/2014CB921203), NSFC (No.11374256/ 11274269), and the
Fundamental Research Funds for the Central Universities in China.

\bibliographystyle{apsrev4-1}
\bibliography{ref}

\begin{thebibliography}{49}%
\makeatletter
\providecommand \@ifxundefined [1]{%
 \@ifx{#1\undefined}
}%
\providecommand \@ifnum [1]{%
 \ifnum #1\expandafter \@firstoftwo
 \else \expandafter \@secondoftwo
 \fi
}%
\providecommand \@ifx [1]{%
 \ifx #1\expandafter \@firstoftwo
 \else \expandafter \@secondoftwo
 \fi
}%
\providecommand \natexlab [1]{#1}%
\providecommand \enquote  [1]{``#1''}%
\providecommand \bibnamefont  [1]{#1}%
\providecommand \bibfnamefont [1]{#1}%
\providecommand \citenamefont [1]{#1}%
\providecommand \href@noop [0]{\@secondoftwo}%
\providecommand \href [0]{\begingroup \@sanitize@url \@href}%
\providecommand \@href[1]{\@@startlink{#1}\@@href}%
\providecommand \@@href[1]{\endgroup#1\@@endlink}%
\providecommand \@sanitize@url [0]{\catcode `\\12\catcode `\$12\catcode
  `\&12\catcode `\#12\catcode `\^12\catcode `\_12\catcode `\%12\relax}%
\providecommand \@@startlink[1]{}%
\providecommand \@@endlink[0]{}%
\providecommand \url  [0]{\begingroup\@sanitize@url \@url }%
\providecommand \@url [1]{\endgroup\@href {#1}{\urlprefix }}%
\providecommand \urlprefix  [0]{URL }%
\providecommand \Eprint [0]{\href }%
\providecommand \doibase [0]{http://dx.doi.org/}%
\providecommand \selectlanguage [0]{\@gobble}%
\providecommand \bibinfo  [0]{\@secondoftwo}%
\providecommand \bibfield  [0]{\@secondoftwo}%
\providecommand \translation [1]{[#1]}%
\providecommand \BibitemOpen [0]{}%
\providecommand \bibitemStop [0]{}%
\providecommand \bibitemNoStop [0]{.\EOS\space}%
\providecommand \EOS [0]{\spacefactor3000\relax}%
\providecommand \BibitemShut  [1]{\csname bibitem#1\endcsname}%
\let\auto@bib@innerbib\@empty
\bibitem [{\citenamefont {Novoselov}\ \emph {et~al.}(2004)\citenamefont
  {Novoselov}, \citenamefont {Geim}, \citenamefont {Morozov}, \citenamefont
  {Jiang}, \citenamefont {Zhang}, \citenamefont {Dubonos}, \citenamefont
  {Grigorieva},\ and\ \citenamefont {Firsov}}]{Graphene2004}%
  \BibitemOpen
  \bibfield  {author} {\bibinfo {author} {\bibfnamefont {K.}~\bibnamefont
  {Novoselov}}, \bibinfo {author} {\bibfnamefont {A.}~\bibnamefont {Geim}},
  \bibinfo {author} {\bibfnamefont {S.}~\bibnamefont {Morozov}}, \bibinfo
  {author} {\bibfnamefont {D.}~\bibnamefont {Jiang}}, \bibinfo {author}
  {\bibfnamefont {Y.}~\bibnamefont {Zhang}}, \bibinfo {author} {\bibfnamefont
  {S.}~\bibnamefont {Dubonos}}, \bibinfo {author} {\bibfnamefont
  {I.}~\bibnamefont {Grigorieva}}, \ and\ \bibinfo {author} {\bibfnamefont
  {A.}~\bibnamefont {Firsov}},\ }\href
  {http://www.scopus.com/inward/record.url?eid=2-s2.0-7444220645&partnerID=40&md5=56c60f8d4125e12ab4579ee5dd194994}
  {\bibfield  {journal} {\bibinfo  {journal} {Science}\ }\textbf {\bibinfo
  {volume} {306}},\ \bibinfo {pages} {666} (\bibinfo {year}
  {2004})}\BibitemShut {NoStop}%
\bibitem [{\citenamefont {Castro~Neto}\ \emph {et~al.}(2009)\citenamefont
  {Castro~Neto}, \citenamefont {Guinea}, \citenamefont {Peres}, \citenamefont
  {Novoselov},\ and\ \citenamefont {Geim}}]{rmp}%
  \BibitemOpen
  \bibfield  {author} {\bibinfo {author} {\bibfnamefont {A.~H.}\ \bibnamefont
  {Castro~Neto}}, \bibinfo {author} {\bibfnamefont {F.}~\bibnamefont {Guinea}},
  \bibinfo {author} {\bibfnamefont {N.~M.~R.}\ \bibnamefont {Peres}}, \bibinfo
  {author} {\bibfnamefont {K.~S.}\ \bibnamefont {Novoselov}}, \ and\ \bibinfo
  {author} {\bibfnamefont {A.~K.}\ \bibnamefont {Geim}},\ }\href {\doibase
  10.1103/RevModPhys.81.109} {\bibfield  {journal} {\bibinfo  {journal} {Rev.
  Mod. Phys.}\ }\textbf {\bibinfo {volume} {81}},\ \bibinfo {pages} {109}
  (\bibinfo {year} {2009})}\BibitemShut {NoStop}%
\bibitem [{\citenamefont {Kotov}\ \emph {et~al.}(2012)\citenamefont {Kotov},
  \citenamefont {Uchoa}, \citenamefont {Pereira}, \citenamefont {Guinea},\ and\
  \citenamefont {Castro~Neto}}]{grapheneinteraction}%
  \BibitemOpen
  \bibfield  {author} {\bibinfo {author} {\bibfnamefont {V.~N.}\ \bibnamefont
  {Kotov}}, \bibinfo {author} {\bibfnamefont {B.}~\bibnamefont {Uchoa}},
  \bibinfo {author} {\bibfnamefont {V.~M.}\ \bibnamefont {Pereira}}, \bibinfo
  {author} {\bibfnamefont {F.}~\bibnamefont {Guinea}}, \ and\ \bibinfo {author}
  {\bibfnamefont {A.~H.}\ \bibnamefont {Castro~Neto}},\ }\href {\doibase
  10.1103/RevModPhys.84.1067} {\bibfield  {journal} {\bibinfo  {journal} {Rev.
  Mod. Phys.}\ }\textbf {\bibinfo {volume} {84}},\ \bibinfo {pages} {1067}
  (\bibinfo {year} {2012})}\BibitemShut {NoStop}%
\bibitem [{\citenamefont {Moore}(2010)}]{Moore2010}%
  \BibitemOpen
  \bibfield  {author} {\bibinfo {author} {\bibfnamefont {J.~E.}\ \bibnamefont
  {Moore}},\ }\href@noop {} {\bibfield  {journal} {\bibinfo  {journal}
  {Nature}\ }\textbf {\bibinfo {volume} {464}},\ \bibinfo {pages} {194}
  (\bibinfo {year} {2010})}\BibitemShut {NoStop}%
\bibitem [{\citenamefont {Hasan}\ and\ \citenamefont {Kane}(2010)}]{Kane2010}%
  \BibitemOpen
  \bibfield  {author} {\bibinfo {author} {\bibfnamefont {M.~Z.}\ \bibnamefont
  {Hasan}}\ and\ \bibinfo {author} {\bibfnamefont {C.~L.}\ \bibnamefont
  {Kane}},\ }\href {\doibase 10.1103/RevModPhys.82.3045} {\bibfield  {journal}
  {\bibinfo  {journal} {Rev. Mod. Phys.}\ }\textbf {\bibinfo {volume} {82}},\
  \bibinfo {pages} {3045} (\bibinfo {year} {2010})}\BibitemShut {NoStop}%
\bibitem [{\citenamefont {Qi}\ and\ \citenamefont {Zhang}(2011)}]{Zhang2011}%
  \BibitemOpen
  \bibfield  {author} {\bibinfo {author} {\bibfnamefont {X.-L.}\ \bibnamefont
  {Qi}}\ and\ \bibinfo {author} {\bibfnamefont {S.-C.}\ \bibnamefont {Zhang}},\
  }\href {\doibase 10.1103/RevModPhys.83.1057} {\bibfield  {journal} {\bibinfo
  {journal} {Rev. Mod. Phys.}\ }\textbf {\bibinfo {volume} {83}},\ \bibinfo
  {pages} {1057} (\bibinfo {year} {2011})}\BibitemShut {NoStop}%
\bibitem [{\citenamefont {Wang}\ \emph {et~al.}(2013)\citenamefont {Wang},
  \citenamefont {Weng}, \citenamefont {Wu}, \citenamefont {Dai},\ and\
  \citenamefont {Fang}}]{Wang2013}%
  \BibitemOpen
  \bibfield  {author} {\bibinfo {author} {\bibfnamefont {Z.}~\bibnamefont
  {Wang}}, \bibinfo {author} {\bibfnamefont {H.}~\bibnamefont {Weng}}, \bibinfo
  {author} {\bibfnamefont {Q.}~\bibnamefont {Wu}}, \bibinfo {author}
  {\bibfnamefont {X.}~\bibnamefont {Dai}}, \ and\ \bibinfo {author}
  {\bibfnamefont {Z.}~\bibnamefont {Fang}},\ }\href {\doibase
  10.1103/PhysRevB.88.125427} {\bibfield  {journal} {\bibinfo  {journal} {Phys.
  Rev. B}\ }\textbf {\bibinfo {volume} {88}},\ \bibinfo {pages} {125427}
  (\bibinfo {year} {2013})}\BibitemShut {NoStop}%
\bibitem [{\citenamefont {Liu}\ \emph {et~al.}(2014{\natexlab{a}})\citenamefont
  {Liu}, \citenamefont {Zhou}, \citenamefont {Zhang}, \citenamefont {Wang},
  \citenamefont {Weng}, \citenamefont {Prabhakaran}, \citenamefont {Mo},
  \citenamefont {Shen}, \citenamefont {Fang}, \citenamefont {Dai} \emph
  {et~al.}}]{liu2014}%
  \BibitemOpen
  \bibfield  {author} {\bibinfo {author} {\bibfnamefont {Z.}~\bibnamefont
  {Liu}}, \bibinfo {author} {\bibfnamefont {B.}~\bibnamefont {Zhou}}, \bibinfo
  {author} {\bibfnamefont {Y.}~\bibnamefont {Zhang}}, \bibinfo {author}
  {\bibfnamefont {Z.}~\bibnamefont {Wang}}, \bibinfo {author} {\bibfnamefont
  {H.}~\bibnamefont {Weng}}, \bibinfo {author} {\bibfnamefont {D.}~\bibnamefont
  {Prabhakaran}}, \bibinfo {author} {\bibfnamefont {S.-K.}\ \bibnamefont {Mo}},
  \bibinfo {author} {\bibfnamefont {Z.}~\bibnamefont {Shen}}, \bibinfo {author}
  {\bibfnamefont {Z.}~\bibnamefont {Fang}}, \bibinfo {author} {\bibfnamefont
  {X.}~\bibnamefont {Dai}},  \emph {et~al.},\ }\href@noop {} {\bibfield
  {journal} {\bibinfo  {journal} {Science}\ }\textbf {\bibinfo {volume}
  {343}},\ \bibinfo {pages} {864} (\bibinfo {year}
  {2014}{\natexlab{a}})}\BibitemShut {NoStop}%
\bibitem [{\citenamefont {Liu}\ \emph {et~al.}(2014{\natexlab{b}})\citenamefont
  {Liu}, \citenamefont {Jiang}, \citenamefont {Zhou}, \citenamefont {Wang},
  \citenamefont {Zhang}, \citenamefont {Weng}, \citenamefont {Prabhakaran},
  \citenamefont {Mo}, \citenamefont {Peng}, \citenamefont {Dudin},
  \citenamefont {Kim}, \citenamefont {Hoesch}, \citenamefont {Fang},
  \citenamefont {Dai}, \citenamefont {Shen}, \citenamefont {Feng},
  \citenamefont {Hussain},\ and\ \citenamefont {Chen}}]{LiuZK2014}%
  \BibitemOpen
  \bibfield  {author} {\bibinfo {author} {\bibfnamefont {Z.}~\bibnamefont
  {Liu}}, \bibinfo {author} {\bibfnamefont {J.}~\bibnamefont {Jiang}}, \bibinfo
  {author} {\bibfnamefont {B.}~\bibnamefont {Zhou}}, \bibinfo {author}
  {\bibfnamefont {Z.}~\bibnamefont {Wang}}, \bibinfo {author} {\bibfnamefont
  {Y.}~\bibnamefont {Zhang}}, \bibinfo {author} {\bibfnamefont
  {H.}~\bibnamefont {Weng}}, \bibinfo {author} {\bibfnamefont {D.}~\bibnamefont
  {Prabhakaran}}, \bibinfo {author} {\bibfnamefont {S.-K.}\ \bibnamefont {Mo}},
  \bibinfo {author} {\bibfnamefont {H.}~\bibnamefont {Peng}}, \bibinfo {author}
  {\bibfnamefont {P.}~\bibnamefont {Dudin}}, \bibinfo {author} {\bibfnamefont
  {T.}~\bibnamefont {Kim}}, \bibinfo {author} {\bibfnamefont {M.}~\bibnamefont
  {Hoesch}}, \bibinfo {author} {\bibfnamefont {Z.}~\bibnamefont {Fang}},
  \bibinfo {author} {\bibfnamefont {X.}~\bibnamefont {Dai}}, \bibinfo {author}
  {\bibfnamefont {Z.}~\bibnamefont {Shen}}, \bibinfo {author} {\bibfnamefont
  {D.}~\bibnamefont {Feng}}, \bibinfo {author} {\bibfnamefont {Z.}~\bibnamefont
  {Hussain}}, \ and\ \bibinfo {author} {\bibfnamefont {Y.}~\bibnamefont
  {Chen}},\ }\href {\doibase 10.1038/nmat3990} {\bibfield  {journal} {\bibinfo
  {journal} {Nature Materials}\ }\textbf {\bibinfo {volume} {13}},\ \bibinfo
  {pages} {677} (\bibinfo {year} {2014}{\natexlab{b}})}\BibitemShut {NoStop}%
\bibitem [{\citenamefont {Neupane}\ \emph {et~al.}(2014)\citenamefont
  {Neupane}, \citenamefont {Xu}, \citenamefont {Sankar}, \citenamefont
  {Alidoust}, \citenamefont {Bian}, \citenamefont {Liu}, \citenamefont
  {Belopolski}, \citenamefont {Chang}, \citenamefont {Jeng}, \citenamefont
  {Lin} \emph {et~al.}}]{neupane2014}%
  \BibitemOpen
  \bibfield  {author} {\bibinfo {author} {\bibfnamefont {M.}~\bibnamefont
  {Neupane}}, \bibinfo {author} {\bibfnamefont {S.-Y.}\ \bibnamefont {Xu}},
  \bibinfo {author} {\bibfnamefont {R.}~\bibnamefont {Sankar}}, \bibinfo
  {author} {\bibfnamefont {N.}~\bibnamefont {Alidoust}}, \bibinfo {author}
  {\bibfnamefont {G.}~\bibnamefont {Bian}}, \bibinfo {author} {\bibfnamefont
  {C.}~\bibnamefont {Liu}}, \bibinfo {author} {\bibfnamefont {I.}~\bibnamefont
  {Belopolski}}, \bibinfo {author} {\bibfnamefont {T.-R.}\ \bibnamefont
  {Chang}}, \bibinfo {author} {\bibfnamefont {H.-T.}\ \bibnamefont {Jeng}},
  \bibinfo {author} {\bibfnamefont {H.}~\bibnamefont {Lin}},  \emph {et~al.},\
  }\href@noop {} {\bibfield  {journal} {\bibinfo  {journal} {Nature
  communications}\ }\textbf {\bibinfo {volume} {5}},\ \bibinfo {pages} {3786}
  (\bibinfo {year} {2014})}\BibitemShut {NoStop}%
\bibitem [{\citenamefont {Wan}\ \emph {et~al.}(2011)\citenamefont {Wan},
  \citenamefont {Turner}, \citenamefont {Vishwanath},\ and\ \citenamefont
  {Savrasov}}]{Wan2011}%
  \BibitemOpen
  \bibfield  {author} {\bibinfo {author} {\bibfnamefont {X.}~\bibnamefont
  {Wan}}, \bibinfo {author} {\bibfnamefont {A.~M.}\ \bibnamefont {Turner}},
  \bibinfo {author} {\bibfnamefont {A.}~\bibnamefont {Vishwanath}}, \ and\
  \bibinfo {author} {\bibfnamefont {S.~Y.}\ \bibnamefont {Savrasov}},\ }\href
  {\doibase 10.1103/PhysRevB.83.205101} {\bibfield  {journal} {\bibinfo
  {journal} {Phys. Rev. B}\ }\textbf {\bibinfo {volume} {83}},\ \bibinfo
  {pages} {205101} (\bibinfo {year} {2011})}\BibitemShut {NoStop}%
\bibitem [{\citenamefont {Burkov}\ \emph {et~al.}(2011)\citenamefont {Burkov},
  \citenamefont {Hook},\ and\ \citenamefont {Balents}}]{Burkov2011}%
  \BibitemOpen
  \bibfield  {author} {\bibinfo {author} {\bibfnamefont {A.~A.}\ \bibnamefont
  {Burkov}}, \bibinfo {author} {\bibfnamefont {M.~D.}\ \bibnamefont {Hook}}, \
  and\ \bibinfo {author} {\bibfnamefont {L.}~\bibnamefont {Balents}},\ }\href
  {\doibase 10.1103/PhysRevB.84.235126} {\bibfield  {journal} {\bibinfo
  {journal} {Phys. Rev. B}\ }\textbf {\bibinfo {volume} {84}},\ \bibinfo
  {pages} {235126} (\bibinfo {year} {2011})}\BibitemShut {NoStop}%
\bibitem [{\citenamefont {Vazifeh}\ and\ \citenamefont
  {Franz}(2013)}]{Vazifeh2013}%
  \BibitemOpen
  \bibfield  {author} {\bibinfo {author} {\bibfnamefont {M.~M.}\ \bibnamefont
  {Vazifeh}}\ and\ \bibinfo {author} {\bibfnamefont {M.}~\bibnamefont
  {Franz}},\ }\href {\doibase 10.1103/PhysRevLett.111.027201} {\bibfield
  {journal} {\bibinfo  {journal} {Phys. Rev. Lett.}\ }\textbf {\bibinfo
  {volume} {111}},\ \bibinfo {pages} {027201} (\bibinfo {year}
  {2013})}\BibitemShut {NoStop}%
\bibitem [{\citenamefont {Weng}\ \emph {et~al.}(2015)\citenamefont {Weng},
  \citenamefont {Fang}, \citenamefont {Fang}, \citenamefont {Bernevig},\ and\
  \citenamefont {Dai}}]{Weng2015}%
  \BibitemOpen
  \bibfield  {author} {\bibinfo {author} {\bibfnamefont {H.}~\bibnamefont
  {Weng}}, \bibinfo {author} {\bibfnamefont {C.}~\bibnamefont {Fang}}, \bibinfo
  {author} {\bibfnamefont {Z.}~\bibnamefont {Fang}}, \bibinfo {author}
  {\bibfnamefont {B.~A.}\ \bibnamefont {Bernevig}}, \ and\ \bibinfo {author}
  {\bibfnamefont {X.}~\bibnamefont {Dai}},\ }\href {\doibase
  10.1103/PhysRevX.5.011029} {\bibfield  {journal} {\bibinfo  {journal} {Phys.
  Rev. X}\ }\textbf {\bibinfo {volume} {5}},\ \bibinfo {pages} {011029}
  (\bibinfo {year} {2015})}\BibitemShut {NoStop}%
\bibitem [{\citenamefont {Huang}\ \emph
  {et~al.}(2015{\natexlab{a}})\citenamefont {Huang}, \citenamefont {Xu},
  \citenamefont {Belopolski}, \citenamefont {Lee}, \citenamefont {Chang},
  \citenamefont {Wang}, \citenamefont {Alidoust}, \citenamefont {Bian},
  \citenamefont {Neupane}, \citenamefont {Zhang}, \citenamefont {Jia},
  \citenamefont {Bansil}, \citenamefont {Lin},\ and\ \citenamefont
  {Hasan}}]{Huang2015}%
  \BibitemOpen
  \bibfield  {author} {\bibinfo {author} {\bibfnamefont {S.-M.}\ \bibnamefont
  {Huang}}, \bibinfo {author} {\bibfnamefont {S.-Y.}\ \bibnamefont {Xu}},
  \bibinfo {author} {\bibfnamefont {I.}~\bibnamefont {Belopolski}}, \bibinfo
  {author} {\bibfnamefont {C.-C.}\ \bibnamefont {Lee}}, \bibinfo {author}
  {\bibfnamefont {G.}~\bibnamefont {Chang}}, \bibinfo {author} {\bibfnamefont
  {B.}~\bibnamefont {Wang}}, \bibinfo {author} {\bibfnamefont {N.}~\bibnamefont
  {Alidoust}}, \bibinfo {author} {\bibfnamefont {G.}~\bibnamefont {Bian}},
  \bibinfo {author} {\bibfnamefont {M.}~\bibnamefont {Neupane}}, \bibinfo
  {author} {\bibfnamefont {C.}~\bibnamefont {Zhang}}, \bibinfo {author}
  {\bibfnamefont {S.}~\bibnamefont {Jia}}, \bibinfo {author} {\bibfnamefont
  {A.}~\bibnamefont {Bansil}}, \bibinfo {author} {\bibfnamefont
  {H.}~\bibnamefont {Lin}}, \ and\ \bibinfo {author} {\bibfnamefont
  {M.}~\bibnamefont {Hasan}},\ }\href
  {http://www.scopus.com/inward/record.url?eid=2-s2.0-84935480287&partnerID=40&md5=b1264d28de2c5c4429fd32bf3f9fefbb}
  {\bibfield  {journal} {\bibinfo  {journal} {Nature Communications}\ }\textbf
  {\bibinfo {volume} {6}},\ \bibinfo {pages} {7373} (\bibinfo {year}
  {2015}{\natexlab{a}})}\BibitemShut {NoStop}%
\bibitem [{\citenamefont {Xu}\ \emph {et~al.}(2015{\natexlab{a}})\citenamefont
  {Xu}, \citenamefont {Belopolski}, \citenamefont {Alidoust}, \citenamefont
  {Neupane}, \citenamefont {Bian}, \citenamefont {Zhang}, \citenamefont
  {Sankar}, \citenamefont {Chang}, \citenamefont {Yuan}, \citenamefont {Lee},
  \citenamefont {Huang}, \citenamefont {Zheng}, \citenamefont {Ma},
  \citenamefont {Sanchez}, \citenamefont {Wang}, \citenamefont {Bansil},
  \citenamefont {Chou}, \citenamefont {Shibayev}, \citenamefont {Lin},
  \citenamefont {Jia},\ and\ \citenamefont {Hasan}}]{Xu2015}%
  \BibitemOpen
  \bibfield  {author} {\bibinfo {author} {\bibfnamefont {S.-Y.}\ \bibnamefont
  {Xu}}, \bibinfo {author} {\bibfnamefont {I.}~\bibnamefont {Belopolski}},
  \bibinfo {author} {\bibfnamefont {N.}~\bibnamefont {Alidoust}}, \bibinfo
  {author} {\bibfnamefont {M.}~\bibnamefont {Neupane}}, \bibinfo {author}
  {\bibfnamefont {G.}~\bibnamefont {Bian}}, \bibinfo {author} {\bibfnamefont
  {C.}~\bibnamefont {Zhang}}, \bibinfo {author} {\bibfnamefont
  {R.}~\bibnamefont {Sankar}}, \bibinfo {author} {\bibfnamefont
  {G.}~\bibnamefont {Chang}}, \bibinfo {author} {\bibfnamefont
  {Z.}~\bibnamefont {Yuan}}, \bibinfo {author} {\bibfnamefont {C.-C.}\
  \bibnamefont {Lee}}, \bibinfo {author} {\bibfnamefont {S.-M.}\ \bibnamefont
  {Huang}}, \bibinfo {author} {\bibfnamefont {H.}~\bibnamefont {Zheng}},
  \bibinfo {author} {\bibfnamefont {J.}~\bibnamefont {Ma}}, \bibinfo {author}
  {\bibfnamefont {D.}~\bibnamefont {Sanchez}}, \bibinfo {author} {\bibfnamefont
  {B.}~\bibnamefont {Wang}}, \bibinfo {author} {\bibfnamefont {A.}~\bibnamefont
  {Bansil}}, \bibinfo {author} {\bibfnamefont {F.}~\bibnamefont {Chou}},
  \bibinfo {author} {\bibfnamefont {P.}~\bibnamefont {Shibayev}}, \bibinfo
  {author} {\bibfnamefont {H.}~\bibnamefont {Lin}}, \bibinfo {author}
  {\bibfnamefont {S.}~\bibnamefont {Jia}}, \ and\ \bibinfo {author}
  {\bibfnamefont {M.}~\bibnamefont {Hasan}},\ }\href {\doibase
  10.1126/science.aaa9297} {\bibfield  {journal} {\bibinfo  {journal}
  {Science}\ }\textbf {\bibinfo {volume} {349}},\ \bibinfo {pages} {613}
  (\bibinfo {year} {2015}{\natexlab{a}})}\BibitemShut {NoStop}%
\bibitem [{\citenamefont {Lv}\ \emph {et~al.}(2015{\natexlab{a}})\citenamefont
  {Lv}, \citenamefont {Weng}, \citenamefont {Fu}, \citenamefont {Wang},
  \citenamefont {Miao}, \citenamefont {Ma}, \citenamefont {Richard},
  \citenamefont {Huang}, \citenamefont {Zhao}, \citenamefont {Chen},
  \citenamefont {Fang}, \citenamefont {Dai}, \citenamefont {Qian},\ and\
  \citenamefont {Ding}}]{Lv2015}%
  \BibitemOpen
  \bibfield  {author} {\bibinfo {author} {\bibfnamefont {B.~Q.}\ \bibnamefont
  {Lv}}, \bibinfo {author} {\bibfnamefont {H.~M.}\ \bibnamefont {Weng}},
  \bibinfo {author} {\bibfnamefont {B.~B.}\ \bibnamefont {Fu}}, \bibinfo
  {author} {\bibfnamefont {X.~P.}\ \bibnamefont {Wang}}, \bibinfo {author}
  {\bibfnamefont {H.}~\bibnamefont {Miao}}, \bibinfo {author} {\bibfnamefont
  {J.}~\bibnamefont {Ma}}, \bibinfo {author} {\bibfnamefont {P.}~\bibnamefont
  {Richard}}, \bibinfo {author} {\bibfnamefont {X.~C.}\ \bibnamefont {Huang}},
  \bibinfo {author} {\bibfnamefont {L.~X.}\ \bibnamefont {Zhao}}, \bibinfo
  {author} {\bibfnamefont {G.~F.}\ \bibnamefont {Chen}}, \bibinfo {author}
  {\bibfnamefont {Z.}~\bibnamefont {Fang}}, \bibinfo {author} {\bibfnamefont
  {X.}~\bibnamefont {Dai}}, \bibinfo {author} {\bibfnamefont {T.}~\bibnamefont
  {Qian}}, \ and\ \bibinfo {author} {\bibfnamefont {H.}~\bibnamefont {Ding}},\
  }\href {\doibase 10.1103/PhysRevX.5.031013} {\bibfield  {journal} {\bibinfo
  {journal} {Phys. Rev. X}\ }\textbf {\bibinfo {volume} {5}},\ \bibinfo {pages}
  {031013} (\bibinfo {year} {2015}{\natexlab{a}})}\BibitemShut {NoStop}%
\bibitem [{\citenamefont {Xu}\ \emph {et~al.}(2015{\natexlab{b}})\citenamefont
  {Xu}, \citenamefont {Alidoust}, \citenamefont {Belopolski}, \citenamefont
  {Yuan}, \citenamefont {Bian}, \citenamefont {Chang}, \citenamefont {Zheng},
  \citenamefont {Strocov}, \citenamefont {Sanchez}, \citenamefont {Chang},
  \citenamefont {Zhang}, \citenamefont {Mou}, \citenamefont {Wu}, \citenamefont
  {Huang}, \citenamefont {Lee}, \citenamefont {Huang}, \citenamefont {Wang},
  \citenamefont {Bansil}, \citenamefont {Jeng}, \citenamefont {Neupert},
  \citenamefont {Kaminski}, \citenamefont {Lin}, \citenamefont {Jia},\ and\
  \citenamefont {Zahid~Hasan}}]{Xu20152}%
  \BibitemOpen
  \bibfield  {author} {\bibinfo {author} {\bibfnamefont {S.-Y.}\ \bibnamefont
  {Xu}}, \bibinfo {author} {\bibfnamefont {N.}~\bibnamefont {Alidoust}},
  \bibinfo {author} {\bibfnamefont {I.}~\bibnamefont {Belopolski}}, \bibinfo
  {author} {\bibfnamefont {Z.}~\bibnamefont {Yuan}}, \bibinfo {author}
  {\bibfnamefont {G.}~\bibnamefont {Bian}}, \bibinfo {author} {\bibfnamefont
  {T.-R.}\ \bibnamefont {Chang}}, \bibinfo {author} {\bibfnamefont
  {H.}~\bibnamefont {Zheng}}, \bibinfo {author} {\bibfnamefont
  {V.}~\bibnamefont {Strocov}}, \bibinfo {author} {\bibfnamefont
  {D.}~\bibnamefont {Sanchez}}, \bibinfo {author} {\bibfnamefont
  {G.}~\bibnamefont {Chang}}, \bibinfo {author} {\bibfnamefont
  {C.}~\bibnamefont {Zhang}}, \bibinfo {author} {\bibfnamefont
  {D.}~\bibnamefont {Mou}}, \bibinfo {author} {\bibfnamefont {Y.}~\bibnamefont
  {Wu}}, \bibinfo {author} {\bibfnamefont {L.}~\bibnamefont {Huang}}, \bibinfo
  {author} {\bibfnamefont {C.-C.}\ \bibnamefont {Lee}}, \bibinfo {author}
  {\bibfnamefont {S.-M.}\ \bibnamefont {Huang}}, \bibinfo {author}
  {\bibfnamefont {B.}~\bibnamefont {Wang}}, \bibinfo {author} {\bibfnamefont
  {A.}~\bibnamefont {Bansil}}, \bibinfo {author} {\bibfnamefont {H.-T.}\
  \bibnamefont {Jeng}}, \bibinfo {author} {\bibfnamefont {T.}~\bibnamefont
  {Neupert}}, \bibinfo {author} {\bibfnamefont {A.}~\bibnamefont {Kaminski}},
  \bibinfo {author} {\bibfnamefont {H.}~\bibnamefont {Lin}}, \bibinfo {author}
  {\bibfnamefont {S.}~\bibnamefont {Jia}}, \ and\ \bibinfo {author}
  {\bibfnamefont {M.}~\bibnamefont {Zahid~Hasan}},\ }\href
  {http://www.nature.com/nphys/journal/v11/n9/full/nphys3437.html} {\bibfield
  {journal} {\bibinfo  {journal} {Nature Physics}\ }\textbf {\bibinfo {volume}
  {11}},\ \bibinfo {pages} {748} (\bibinfo {year}
  {2015}{\natexlab{b}})}\BibitemShut {NoStop}%
\bibitem [{\citenamefont {Zhang}\ \emph {et~al.}(2015)\citenamefont {Zhang},
  \citenamefont {Yuan}, \citenamefont {Xu}, \citenamefont {Lin}, \citenamefont
  {Tong}, \citenamefont {Hasan}, \citenamefont {Wang}, \citenamefont {Zhang},\
  and\ \citenamefont {Jia}}]{Zhang2015}%
  \BibitemOpen
  \bibfield  {author} {\bibinfo {author} {\bibfnamefont {C.}~\bibnamefont
  {Zhang}}, \bibinfo {author} {\bibfnamefont {Z.}~\bibnamefont {Yuan}},
  \bibinfo {author} {\bibfnamefont {S.}~\bibnamefont {Xu}}, \bibinfo {author}
  {\bibfnamefont {Z.}~\bibnamefont {Lin}}, \bibinfo {author} {\bibfnamefont
  {B.}~\bibnamefont {Tong}}, \bibinfo {author} {\bibfnamefont {M.~Z.}\
  \bibnamefont {Hasan}}, \bibinfo {author} {\bibfnamefont {J.}~\bibnamefont
  {Wang}}, \bibinfo {author} {\bibfnamefont {C.}~\bibnamefont {Zhang}}, \ and\
  \bibinfo {author} {\bibfnamefont {S.}~\bibnamefont {Jia}},\ }\href@noop {}
  {\bibfield  {journal} {\bibinfo  {journal} {arXiv preprint arXiv:1502.00251}\
  } (\bibinfo {year} {2015})}\BibitemShut {NoStop}%
\bibitem [{\citenamefont {Yang}\ \emph {et~al.}(2015)\citenamefont {Yang},
  \citenamefont {Liu}, \citenamefont {Sun}, \citenamefont {Peng}, \citenamefont
  {Yang}, \citenamefont {Zhang}, \citenamefont {Zhou}, \citenamefont {Zhang},
  \citenamefont {Guo}, \citenamefont {Rahn}, \citenamefont {Prabhakaran},
  \citenamefont {Hussain}, \citenamefont {Mo}, \citenamefont {Felser},
  \citenamefont {Yan},\ and\ \citenamefont {Chen}}]{Yang2015}%
  \BibitemOpen
  \bibfield  {author} {\bibinfo {author} {\bibfnamefont {L.}~\bibnamefont
  {Yang}}, \bibinfo {author} {\bibfnamefont {Z.}~\bibnamefont {Liu}}, \bibinfo
  {author} {\bibfnamefont {Y.}~\bibnamefont {Sun}}, \bibinfo {author}
  {\bibfnamefont {H.}~\bibnamefont {Peng}}, \bibinfo {author} {\bibfnamefont
  {H.}~\bibnamefont {Yang}}, \bibinfo {author} {\bibfnamefont {T.}~\bibnamefont
  {Zhang}}, \bibinfo {author} {\bibfnamefont {B.}~\bibnamefont {Zhou}},
  \bibinfo {author} {\bibfnamefont {Y.}~\bibnamefont {Zhang}}, \bibinfo
  {author} {\bibfnamefont {Y.}~\bibnamefont {Guo}}, \bibinfo {author}
  {\bibfnamefont {M.}~\bibnamefont {Rahn}}, \bibinfo {author} {\bibfnamefont
  {D.}~\bibnamefont {Prabhakaran}}, \bibinfo {author} {\bibfnamefont
  {Z.}~\bibnamefont {Hussain}}, \bibinfo {author} {\bibfnamefont {S.-K.}\
  \bibnamefont {Mo}}, \bibinfo {author} {\bibfnamefont {C.}~\bibnamefont
  {Felser}}, \bibinfo {author} {\bibfnamefont {B.}~\bibnamefont {Yan}}, \ and\
  \bibinfo {author} {\bibfnamefont {Y.}~\bibnamefont {Chen}},\ }\href
  {http://www.nature.com/nphys/journal/v11/n9/full/nphys3425.html} {\bibfield
  {journal} {\bibinfo  {journal} {Nature Physics}\ }\textbf {\bibinfo {volume}
  {11}},\ \bibinfo {pages} {728} (\bibinfo {year} {2015})}\BibitemShut
  {NoStop}%
\bibitem [{\citenamefont {Wang}\ \emph {et~al.}(2015)\citenamefont {Wang},
  \citenamefont {Zheng}, \citenamefont {Shen}, \citenamefont {Zhou},
  \citenamefont {Yang}, \citenamefont {Li}, \citenamefont {Feng},\ and\
  \citenamefont {Xu}}]{Wang2015}%
  \BibitemOpen
  \bibfield  {author} {\bibinfo {author} {\bibfnamefont {Z.}~\bibnamefont
  {Wang}}, \bibinfo {author} {\bibfnamefont {Y.}~\bibnamefont {Zheng}},
  \bibinfo {author} {\bibfnamefont {Z.}~\bibnamefont {Shen}}, \bibinfo {author}
  {\bibfnamefont {Y.}~\bibnamefont {Zhou}}, \bibinfo {author} {\bibfnamefont
  {X.}~\bibnamefont {Yang}}, \bibinfo {author} {\bibfnamefont {Y.}~\bibnamefont
  {Li}}, \bibinfo {author} {\bibfnamefont {C.}~\bibnamefont {Feng}}, \ and\
  \bibinfo {author} {\bibfnamefont {Z.-A.}\ \bibnamefont {Xu}},\ }\href@noop {}
  {\bibfield  {journal} {\bibinfo  {journal} {arXiv preprint arXiv:1506.00924}\
  } (\bibinfo {year} {2015})}\BibitemShut {NoStop}%
\bibitem [{\citenamefont {Huang}\ \emph
  {et~al.}(2015{\natexlab{b}})\citenamefont {Huang}, \citenamefont {Zhao},
  \citenamefont {Long}, \citenamefont {Wang}, \citenamefont {Chen},
  \citenamefont {Yang}, \citenamefont {Liang}, \citenamefont {Xue},
  \citenamefont {Weng}, \citenamefont {Fang}, \citenamefont {Dai},\ and\
  \citenamefont {Chen}}]{HuangXC2015}%
  \BibitemOpen
  \bibfield  {author} {\bibinfo {author} {\bibfnamefont {X.}~\bibnamefont
  {Huang}}, \bibinfo {author} {\bibfnamefont {L.}~\bibnamefont {Zhao}},
  \bibinfo {author} {\bibfnamefont {Y.}~\bibnamefont {Long}}, \bibinfo {author}
  {\bibfnamefont {P.}~\bibnamefont {Wang}}, \bibinfo {author} {\bibfnamefont
  {D.}~\bibnamefont {Chen}}, \bibinfo {author} {\bibfnamefont {Z.}~\bibnamefont
  {Yang}}, \bibinfo {author} {\bibfnamefont {H.}~\bibnamefont {Liang}},
  \bibinfo {author} {\bibfnamefont {M.}~\bibnamefont {Xue}}, \bibinfo {author}
  {\bibfnamefont {H.}~\bibnamefont {Weng}}, \bibinfo {author} {\bibfnamefont
  {Z.}~\bibnamefont {Fang}}, \bibinfo {author} {\bibfnamefont {X.}~\bibnamefont
  {Dai}}, \ and\ \bibinfo {author} {\bibfnamefont {G.}~\bibnamefont {Chen}},\
  }\href {\doibase 10.1103/PhysRevX.5.031023} {\bibfield  {journal} {\bibinfo
  {journal} {Phys. Rev. X}\ }\textbf {\bibinfo {volume} {5}},\ \bibinfo {pages}
  {031023} (\bibinfo {year} {2015}{\natexlab{b}})}\BibitemShut {NoStop}%
\bibitem [{\citenamefont {Lv}\ \emph {et~al.}(2015{\natexlab{b}})\citenamefont
  {Lv}, \citenamefont {Xu}, \citenamefont {Weng}, \citenamefont {Ma},
  \citenamefont {Richard}, \citenamefont {Huang}, \citenamefont {Zhao},
  \citenamefont {Chen}, \citenamefont {Matt}, \citenamefont {Bisti},
  \citenamefont {Strocov}, \citenamefont {Mesot}, \citenamefont {Fang},
  \citenamefont {Dai}, \citenamefont {Qian}, \citenamefont {Shi},\ and\
  \citenamefont {Ding}}]{Lv2015724}%
  \BibitemOpen
  \bibfield  {author} {\bibinfo {author} {\bibfnamefont {B.}~\bibnamefont
  {Lv}}, \bibinfo {author} {\bibfnamefont {N.}~\bibnamefont {Xu}}, \bibinfo
  {author} {\bibfnamefont {H.}~\bibnamefont {Weng}}, \bibinfo {author}
  {\bibfnamefont {J.}~\bibnamefont {Ma}}, \bibinfo {author} {\bibfnamefont
  {P.}~\bibnamefont {Richard}}, \bibinfo {author} {\bibfnamefont
  {X.}~\bibnamefont {Huang}}, \bibinfo {author} {\bibfnamefont
  {L.}~\bibnamefont {Zhao}}, \bibinfo {author} {\bibfnamefont {G.}~\bibnamefont
  {Chen}}, \bibinfo {author} {\bibfnamefont {C.}~\bibnamefont {Matt}}, \bibinfo
  {author} {\bibfnamefont {F.}~\bibnamefont {Bisti}}, \bibinfo {author}
  {\bibfnamefont {V.}~\bibnamefont {Strocov}}, \bibinfo {author} {\bibfnamefont
  {J.}~\bibnamefont {Mesot}}, \bibinfo {author} {\bibfnamefont
  {Z.}~\bibnamefont {Fang}}, \bibinfo {author} {\bibfnamefont {X.}~\bibnamefont
  {Dai}}, \bibinfo {author} {\bibfnamefont {T.}~\bibnamefont {Qian}}, \bibinfo
  {author} {\bibfnamefont {M.}~\bibnamefont {Shi}}, \ and\ \bibinfo {author}
  {\bibfnamefont {H.}~\bibnamefont {Ding}},\ }\href {\doibase
  10.1038/nphys3426} {\bibfield  {journal} {\bibinfo  {journal} {Nature
  Physics}\ }\textbf {\bibinfo {volume} {11}},\ \bibinfo {pages} {724}
  (\bibinfo {year} {2015}{\natexlab{b}})}\BibitemShut {NoStop}%
\bibitem [{\citenamefont {Anderson}(1961)}]{anderson1961}%
  \BibitemOpen
  \bibfield  {author} {\bibinfo {author} {\bibfnamefont {P.~W.}\ \bibnamefont
  {Anderson}},\ }\href@noop {} {\bibfield  {journal} {\bibinfo  {journal}
  {Physical Review}\ }\textbf {\bibinfo {volume} {124}},\ \bibinfo {pages} {41}
  (\bibinfo {year} {1961})}\BibitemShut {NoStop}%
\bibitem [{\citenamefont {Kondo}(1964)}]{kondo1964}%
  \BibitemOpen
  \bibfield  {author} {\bibinfo {author} {\bibfnamefont {J.}~\bibnamefont
  {Kondo}},\ }\href@noop {} {\bibfield  {journal} {\bibinfo  {journal}
  {Progress of theoretical physics}\ }\textbf {\bibinfo {volume} {32}},\
  \bibinfo {pages} {37} (\bibinfo {year} {1964})}\BibitemShut {NoStop}%
\bibitem [{\citenamefont {Krishna-murthy}\ \emph {et~al.}(1980)\citenamefont
  {Krishna-murthy}, \citenamefont {Wilkins},\ and\ \citenamefont
  {Wilson}}]{Krishna1980}%
  \BibitemOpen
  \bibfield  {author} {\bibinfo {author} {\bibfnamefont {H.~R.}\ \bibnamefont
  {Krishna-murthy}}, \bibinfo {author} {\bibfnamefont {J.~W.}\ \bibnamefont
  {Wilkins}}, \ and\ \bibinfo {author} {\bibfnamefont {K.~G.}\ \bibnamefont
  {Wilson}},\ }\href {\doibase 10.1103/PhysRevB.21.1003} {\bibfield  {journal}
  {\bibinfo  {journal} {Phys. Rev. B}\ }\textbf {\bibinfo {volume} {21}},\
  \bibinfo {pages} {1003} (\bibinfo {year} {1980})}\BibitemShut {NoStop}%
\bibitem [{\citenamefont {Tsvelick}\ and\ \citenamefont
  {Wiegmann}(1984)}]{tsvelick1984}%
  \BibitemOpen
  \bibfield  {author} {\bibinfo {author} {\bibfnamefont {A.}~\bibnamefont
  {Tsvelick}}\ and\ \bibinfo {author} {\bibfnamefont {P.}~\bibnamefont
  {Wiegmann}},\ }\href@noop {} {\bibfield  {journal} {\bibinfo  {journal}
  {Zeitschrift f{\"u}r Physik B Condensed Matter}\ }\textbf {\bibinfo {volume}
  {54}},\ \bibinfo {pages} {201} (\bibinfo {year} {1984})}\BibitemShut
  {NoStop}%
\bibitem [{\citenamefont {Andrei}\ and\ \citenamefont
  {Destri}(1984)}]{Andrei1984}%
  \BibitemOpen
  \bibfield  {author} {\bibinfo {author} {\bibfnamefont {N.}~\bibnamefont
  {Andrei}}\ and\ \bibinfo {author} {\bibfnamefont {C.}~\bibnamefont
  {Destri}},\ }\href {\doibase 10.1103/PhysRevLett.52.364} {\bibfield
  {journal} {\bibinfo  {journal} {Phys. Rev. Lett.}\ }\textbf {\bibinfo
  {volume} {52}},\ \bibinfo {pages} {364} (\bibinfo {year} {1984})}\BibitemShut
  {NoStop}%
\bibitem [{\citenamefont {Zhang}\ and\ \citenamefont {Lee}(1983)}]{Zhang1983}%
  \BibitemOpen
  \bibfield  {author} {\bibinfo {author} {\bibfnamefont {F.~C.}\ \bibnamefont
  {Zhang}}\ and\ \bibinfo {author} {\bibfnamefont {T.~K.}\ \bibnamefont
  {Lee}},\ }\href {\doibase 10.1103/PhysRevB.28.33} {\bibfield  {journal}
  {\bibinfo  {journal} {Phys. Rev. B}\ }\textbf {\bibinfo {volume} {28}},\
  \bibinfo {pages} {33} (\bibinfo {year} {1983})}\BibitemShut {NoStop}%
\bibitem [{\citenamefont {Coleman}(1984)}]{Coleman1984}%
  \BibitemOpen
  \bibfield  {author} {\bibinfo {author} {\bibfnamefont {P.}~\bibnamefont
  {Coleman}},\ }\href {\doibase 10.1103/PhysRevB.29.3035} {\bibfield  {journal}
  {\bibinfo  {journal} {Phys. Rev. B}\ }\textbf {\bibinfo {volume} {29}},\
  \bibinfo {pages} {3035} (\bibinfo {year} {1984})}\BibitemShut {NoStop}%
\bibitem [{\citenamefont {Read}\ and\ \citenamefont {Newns}(1983)}]{read1983}%
  \BibitemOpen
  \bibfield  {author} {\bibinfo {author} {\bibfnamefont {N.}~\bibnamefont
  {Read}}\ and\ \bibinfo {author} {\bibfnamefont {D.}~\bibnamefont {Newns}},\
  }\href@noop {} {\bibfield  {journal} {\bibinfo  {journal} {Journal of Physics
  C: Solid State Physics}\ }\textbf {\bibinfo {volume} {16}},\ \bibinfo {pages}
  {3273} (\bibinfo {year} {1983})}\BibitemShut {NoStop}%
\bibitem [{\citenamefont {Kuramoto}(1983)}]{Kuramoto1983}%
  \BibitemOpen
  \bibfield  {author} {\bibinfo {author} {\bibfnamefont {Y.}~\bibnamefont
  {Kuramoto}},\ }\href {\doibase 10.1007/BF01578246} {\bibfield  {journal}
  {\bibinfo  {journal} {Zeitschrift f$\ddot{u}$r Physik B Condensed Matter}\
  }\textbf {\bibinfo {volume} {53}},\ \bibinfo {pages} {37} (\bibinfo {year}
  {1983})}\BibitemShut {NoStop}%
\bibitem [{\citenamefont {Gunnarsson}\ and\ \citenamefont
  {Sch\"onhammer}(1983)}]{Gunnarsson1983}%
  \BibitemOpen
  \bibfield  {author} {\bibinfo {author} {\bibfnamefont {O.}~\bibnamefont
  {Gunnarsson}}\ and\ \bibinfo {author} {\bibfnamefont {K.}~\bibnamefont
  {Sch\"onhammer}},\ }\href {\doibase 10.1103/PhysRevLett.50.604} {\bibfield
  {journal} {\bibinfo  {journal} {Phys. Rev. Lett.}\ }\textbf {\bibinfo
  {volume} {50}},\ \bibinfo {pages} {604} (\bibinfo {year} {1983})}\BibitemShut
  {NoStop}%
\bibitem [{\citenamefont {Affleck}(1990)}]{affleck1990}%
  \BibitemOpen
  \bibfield  {author} {\bibinfo {author} {\bibfnamefont {I.}~\bibnamefont
  {Affleck}},\ }\href@noop {} {\bibfield  {journal} {\bibinfo  {journal}
  {Nuclear Physics B}\ }\textbf {\bibinfo {volume} {336}},\ \bibinfo {pages}
  {517} (\bibinfo {year} {1990})}\BibitemShut {NoStop}%
\bibitem [{\citenamefont {Ishii}(1978)}]{ishii1978}%
  \BibitemOpen
  \bibfield  {author} {\bibinfo {author} {\bibfnamefont {H.}~\bibnamefont
  {Ishii}},\ }\href@noop {} {\bibfield  {journal} {\bibinfo  {journal} {Journal
  of Low Temperature Physics}\ }\textbf {\bibinfo {volume} {32}},\ \bibinfo
  {pages} {457} (\bibinfo {year} {1978})}\BibitemShut {NoStop}%
\bibitem [{\citenamefont {Barzykin}\ and\ \citenamefont
  {Affleck}(1998)}]{Barzykin1998}%
  \BibitemOpen
  \bibfield  {author} {\bibinfo {author} {\bibfnamefont {V.}~\bibnamefont
  {Barzykin}}\ and\ \bibinfo {author} {\bibfnamefont {I.}~\bibnamefont
  {Affleck}},\ }\href {\doibase 10.1103/PhysRevB.57.432} {\bibfield  {journal}
  {\bibinfo  {journal} {Phys. Rev. B}\ }\textbf {\bibinfo {volume} {57}},\
  \bibinfo {pages} {432} (\bibinfo {year} {1998})}\BibitemShut {NoStop}%
\bibitem [{\citenamefont {Borda}(2007)}]{Borda2007}%
  \BibitemOpen
  \bibfield  {author} {\bibinfo {author} {\bibfnamefont {L.}~\bibnamefont
  {Borda}},\ }\href {\doibase 10.1103/PhysRevB.75.041307} {\bibfield  {journal}
  {\bibinfo  {journal} {Phys. Rev. B}\ }\textbf {\bibinfo {volume} {75}},\
  \bibinfo {pages} {041307} (\bibinfo {year} {2007})}\BibitemShut {NoStop}%
\bibitem [{\citenamefont {Gonzalez-Buxton}\ and\ \citenamefont
  {Ingersent}(1998)}]{Gonzalez1998}%
  \BibitemOpen
  \bibfield  {author} {\bibinfo {author} {\bibfnamefont {C.}~\bibnamefont
  {Gonzalez-Buxton}}\ and\ \bibinfo {author} {\bibfnamefont {K.}~\bibnamefont
  {Ingersent}},\ }\href {\doibase 10.1103/PhysRevB.57.14254} {\bibfield
  {journal} {\bibinfo  {journal} {Phys. Rev. B}\ }\textbf {\bibinfo {volume}
  {57}},\ \bibinfo {pages} {14254} (\bibinfo {year} {1998})}\BibitemShut
  {NoStop}%
\bibitem [{\citenamefont {Fritz}\ and\ \citenamefont
  {Vojta}(2004)}]{Fritz2004}%
  \BibitemOpen
  \bibfield  {author} {\bibinfo {author} {\bibfnamefont {L.}~\bibnamefont
  {Fritz}}\ and\ \bibinfo {author} {\bibfnamefont {M.}~\bibnamefont {Vojta}},\
  }\href {\doibase 10.1103/PhysRevB.70.214427} {\bibfield  {journal} {\bibinfo
  {journal} {Phys. Rev. B}\ }\textbf {\bibinfo {volume} {70}},\ \bibinfo
  {pages} {214427} (\bibinfo {year} {2004})}\BibitemShut {NoStop}%
\bibitem [{\citenamefont {Vojta}\ and\ \citenamefont
  {Fritz}(2004)}]{Vojta2004}%
  \BibitemOpen
  \bibfield  {author} {\bibinfo {author} {\bibfnamefont {M.}~\bibnamefont
  {Vojta}}\ and\ \bibinfo {author} {\bibfnamefont {L.}~\bibnamefont {Fritz}},\
  }\href {\doibase 10.1103/PhysRevB.70.094502} {\bibfield  {journal} {\bibinfo
  {journal} {Phys. Rev. B}\ }\textbf {\bibinfo {volume} {70}},\ \bibinfo
  {pages} {094502} (\bibinfo {year} {2004})}\BibitemShut {NoStop}%
\bibitem [{\citenamefont {Shirakawa}\ and\ \citenamefont
  {Yunoki}(2014)}]{shirakawa2014}%
  \BibitemOpen
  \bibfield  {author} {\bibinfo {author} {\bibfnamefont {T.}~\bibnamefont
  {Shirakawa}}\ and\ \bibinfo {author} {\bibfnamefont {S.}~\bibnamefont
  {Yunoki}},\ }\href {\doibase 10.1103/PhysRevB.90.195109} {\bibfield
  {journal} {\bibinfo  {journal} {Phys. Rev. B}\ }\textbf {\bibinfo {volume}
  {90}},\ \bibinfo {pages} {195109} (\bibinfo {year} {2014})}\BibitemShut
  {NoStop}%
\bibitem [{\citenamefont {Mitchell}\ and\ \citenamefont
  {Fritz}(2015)}]{Mitchell2015}%
  \BibitemOpen
  \bibfield  {author} {\bibinfo {author} {\bibfnamefont {A.~K.}\ \bibnamefont
  {Mitchell}}\ and\ \bibinfo {author} {\bibfnamefont {L.}~\bibnamefont
  {Fritz}},\ }\href {\doibase 10.1103/PhysRevB.92.121109} {\bibfield  {journal}
  {\bibinfo  {journal} {Phys. Rev. B}\ }\textbf {\bibinfo {volume} {92}},\
  \bibinfo {pages} {121109} (\bibinfo {year} {2015})}\BibitemShut {NoStop}%
\bibitem [{\citenamefont {Varma}\ and\ \citenamefont
  {Yafet}(1976)}]{Varma1976}%
  \BibitemOpen
  \bibfield  {author} {\bibinfo {author} {\bibfnamefont {C.~M.}\ \bibnamefont
  {Varma}}\ and\ \bibinfo {author} {\bibfnamefont {Y.}~\bibnamefont {Yafet}},\
  }\href {\doibase 10.1103/PhysRevB.13.2950} {\bibfield  {journal} {\bibinfo
  {journal} {Phys. Rev. B}\ }\textbf {\bibinfo {volume} {13}},\ \bibinfo
  {pages} {2950} (\bibinfo {year} {1976})}\BibitemShut {NoStop}%
\bibitem [{\citenamefont {Aji}\ \emph {et~al.}(2008)\citenamefont {Aji},
  \citenamefont {Varma},\ and\ \citenamefont {Vekhter}}]{Aji2008}%
  \BibitemOpen
  \bibfield  {author} {\bibinfo {author} {\bibfnamefont {V.}~\bibnamefont
  {Aji}}, \bibinfo {author} {\bibfnamefont {C.~M.}\ \bibnamefont {Varma}}, \
  and\ \bibinfo {author} {\bibfnamefont {I.}~\bibnamefont {Vekhter}},\ }\href
  {\doibase 10.1103/PhysRevB.77.224426} {\bibfield  {journal} {\bibinfo
  {journal} {Phys. Rev. B}\ }\textbf {\bibinfo {volume} {77}},\ \bibinfo
  {pages} {224426} (\bibinfo {year} {2008})}\BibitemShut {NoStop}%
\bibitem [{\citenamefont {Feng}\ \emph {et~al.}(2010)\citenamefont {Feng},
  \citenamefont {Chen}, \citenamefont {Gao}, \citenamefont {Wang},\ and\
  \citenamefont {Zhang}}]{Feng2010}%
  \BibitemOpen
  \bibfield  {author} {\bibinfo {author} {\bibfnamefont {X.-Y.}\ \bibnamefont
  {Feng}}, \bibinfo {author} {\bibfnamefont {W.-Q.}\ \bibnamefont {Chen}},
  \bibinfo {author} {\bibfnamefont {J.-H.}\ \bibnamefont {Gao}}, \bibinfo
  {author} {\bibfnamefont {Q.-H.}\ \bibnamefont {Wang}}, \ and\ \bibinfo
  {author} {\bibfnamefont {F.-C.}\ \bibnamefont {Zhang}},\ }\href {\doibase
  10.1103/PhysRevB.81.235411} {\bibfield  {journal} {\bibinfo  {journal} {Phys.
  Rev. B}\ }\textbf {\bibinfo {volume} {81}},\ \bibinfo {pages} {235411}
  (\bibinfo {year} {2010})}\BibitemShut {NoStop}%
\bibitem [{\citenamefont {Fu}\ and\ \citenamefont {Berg}(2010)}]{FuLiang2010}%
  \BibitemOpen
  \bibfield  {author} {\bibinfo {author} {\bibfnamefont {L.}~\bibnamefont
  {Fu}}\ and\ \bibinfo {author} {\bibfnamefont {E.}~\bibnamefont {Berg}},\
  }\href {\doibase 10.1103/PhysRevLett.105.097001} {\bibfield  {journal}
  {\bibinfo  {journal} {Phys. Rev. Lett.}\ }\textbf {\bibinfo {volume} {105}},\
  \bibinfo {pages} {097001} (\bibinfo {year} {2010})}\BibitemShut {NoStop}%
\bibitem [{\citenamefont {Sengupta}\ and\ \citenamefont
  {Baskaran}(2008)}]{Sengupta2008}%
  \BibitemOpen
  \bibfield  {author} {\bibinfo {author} {\bibfnamefont {K.}~\bibnamefont
  {Sengupta}}\ and\ \bibinfo {author} {\bibfnamefont {G.}~\bibnamefont
  {Baskaran}},\ }\href {\doibase 10.1103/PhysRevB.77.045417} {\bibfield
  {journal} {\bibinfo  {journal} {Phys. Rev. B}\ }\textbf {\bibinfo {volume}
  {77}},\ \bibinfo {pages} {045417} (\bibinfo {year} {2008})}\BibitemShut
  {NoStop}%
\bibitem [{\citenamefont {Zhuang}\ \emph {et~al.}(2009)\citenamefont {Zhuang},
  \citenamefont {Sun},\ and\ \citenamefont {Xie}}]{Zhuang2009}%
  \BibitemOpen
  \bibfield  {author} {\bibinfo {author} {\bibfnamefont {H.-B.}\ \bibnamefont
  {Zhuang}}, \bibinfo {author} {\bibfnamefont {Q.-F.}\ \bibnamefont {Sun}}, \
  and\ \bibinfo {author} {\bibfnamefont {X.}~\bibnamefont {Xie}},\ }\href
  {http://iopscience.iop.org/article/10.1209/0295-5075/86/58004/pdf} {\bibfield
   {journal} {\bibinfo  {journal} {EPL}\ }\textbf {\bibinfo {volume} {86}},\
  \bibinfo {pages} {58004} (\bibinfo {year} {2009})}\BibitemShut {NoStop}%
\bibitem [{\citenamefont {Principi}\ \emph {et~al.}(2015)\citenamefont
  {Principi}, \citenamefont {Vignale},\ and\ \citenamefont
  {Rossi}}]{Rossi2014}%
  \BibitemOpen
  \bibfield  {author} {\bibinfo {author} {\bibfnamefont {A.}~\bibnamefont
  {Principi}}, \bibinfo {author} {\bibfnamefont {G.}~\bibnamefont {Vignale}}, \
  and\ \bibinfo {author} {\bibfnamefont {E.}~\bibnamefont {Rossi}},\ }\href
  {\doibase 10.1103/PhysRevB.92.041107} {\bibfield  {journal} {\bibinfo
  {journal} {Phys. Rev. B}\ }\textbf {\bibinfo {volume} {92}},\ \bibinfo
  {pages} {041107} (\bibinfo {year} {2015})}\BibitemShut {NoStop}%
\end{thebibliography}%

\end{document}